\documentclass[prd,aps,showpacs,tightenlines,reprint,nofootinbib,superscriptaddress,floatfix,twocolumn]{revtex4-2}
\usepackage{epsfig}
\usepackage{amsfonts}
\usepackage{amsmath}
\usepackage{bbm,bm}
\usepackage{xcolor}
\usepackage{graphicx}
\usepackage{enumerate}

\usepackage[
  breaklinks,
  pdfencoding=auto,
  psdextra,
  colorlinks,
  linkcolor={red!50!black},
  citecolor={blue!50!black},
  urlcolor={blue!80!black}
]{hyperref}
\usepackage[normalem]{ulem}

\newcommand{\nn}{\nonumber \\}
\newcommand{\rmi}[1]{{\mbox{\scriptsize #1}}}
\newcommand{\rmii}[1]{{\mbox{\tiny\rm{#1}}}}

\newcommand{\msl}[1]{\,\slash\!\!\!{#1}\,}
\newcommand{\tr}{{\rm Tr}}
\renewcommand{\vec}[1]{{\bf #1}}
\newcommand{\T}{\rmii{$T$}}
\newcommand{\glue}{\rmi{$\mathrm{glue}$}}
\newcommand{\gs}{g_\rmi{s}}
\newcommand{\gbl}{g_{\rmii{$B$-$L$}}}

\newcommand{\csb}{\chi\mathrm{SB}}
\newcommand{\SU}{{\rm SU}}
\newcommand{\UBL}{{\rm U}(1)_{\rmii{$B$-$L$}}}
\newcommand{\SUX}{{\rm SU}(2)_{\rmii{$X$}}}
\newcommand{\vphi}{v_\Phi}
\newcommand{\vhqcd}{v_{h,\rmii{QCD}}}

\newcommand{\lp}{\lambda_p}
\newcommand{\lh}{\lambda_H}
\newcommand{\lphi}{\lambda_\Phi}

\newcommand{\MeV}{{\rm MeV}}
\newcommand{\GeV}{{\rm GeV}}

\newcommand{\mW}{m_\rmii{$W$}}
\newcommand{\mZ}{m_\rmii{$Z$}}
\newcommand{\MPl}{M_\rmii{Pl}}

\newcommand{\Veff}{V_\mathrm{eff}}
\newcommand{\VeffNJL}{V_\mathrm{eff}^\rmii{NJL}}
\newcommand{\VNJL}{V^\rmii{NJL}}
\newcommand{\VeffPNJL}{V_\mathrm{eff}^\rmii{PNJL}}
\newcommand{\VPNJL}{V^\rmii{PNJL}}

\newcommand{\vw}{v_\rmi{w}}
\newcommand{\gammaeq}{\gamma_\mathrm{eq}}

\newcommand{\Tc}{T_{\rm c}}
\newcommand{\Tn}{T_{\rm n}}
\newcommand{\Tp}{T_p}
\newcommand{\Ti}{T_i}
\newcommand{\Nf}{N_{\rm f}}
\newcommand{\Nc}{N_{\rm c}}

\newcommand{\TRH}{T_\rmii{RH}}
\newcommand{\TEW}{T_\rmii{EW}}
\newcommand{\TQCD}{T_\rmii{QCD}}

\newcommand{\Tint}[1]{{\hbox{$\sum$}\!\!\!\!\!\!\!\int\,}_{\!\!\!\!\raise-0.9ex\hbox{$\scriptstyle{#1}$}}}
\newcommand{\Tinti}[1]{{{\Sigma}\!\!\!\!\raise0.3ex\hbox{$\int$}_\rmii{${#1}$}}}
\newcommand{\Tintii}[2]{{{\Sigma^\rmii{${#2}$}}\!\!\!\!\!\!\!\raise0.3ex\hbox{$\int$}_\rmii{${#1}$}}}
\newcommand{\Tintip}[1]{{{\Sigma'}\!\!\!\!\!\raise0.3ex\hbox{$\int$}_\rmii{${#1}$}}}

\begin{document}

%%%%%%%%%%%%%%%%%%%%%%%% Title page %%%%%
\title{Supercool exit: Gravitational waves from \\QCD-triggered conformal symmetry breaking}

\author{Laura Sagunski}
\email{sagunski@itp.uni-frankfurt.de}
\affiliation{Institute for Theoretical Physics, Goethe University, 60438 Frankfurt am Main, Germany}

\author{Philipp Schicho}
\email{schicho@itp.uni-frankfurt.de}
\affiliation{Institute for Theoretical Physics, Goethe University, 60438 Frankfurt am Main, Germany}

\author{Daniel Schmitt}
\email{dschmitt@itp.uni-frankfurt.de}
\affiliation{Institute for Theoretical Physics, Goethe University, 60438 Frankfurt am Main, Germany}

%\date{\today}

\begin{abstract}
Classically conformal Standard Model extensions predict an intriguing thermal history of the early Universe.
In contrast to the common paradigm, the onset of the electroweak phase transition can be significantly delayed while
the Universe undergoes a period of thermal inflation.
Then, a first-order chiral phase transition could not only
trigger electroweak symmetry breaking but also
initiate the exit from supercooling.
To study the dynamics of this scenario,
we focus on
low-energy quark-based QCD effective models that exhibit a first-order transition.
While a large amount of latent heat is naturally involved if thermal inflation ends,
we find that a supercooling period prior to the QCD scale considerably enhances the timescale of the transition.
This enhancement implies great observational prospects at future gravitational wave observatories.
Our results are readily applicable to a wide class of scale-invariant SM extensions, as well as strongly coupled dark sectors.
\end{abstract}
%\preprint{}
%\pacs{gravitational waves, dark matter, \SRC{pacs are usually alphanumerical codes}}

\maketitle

%%%%%%%%%%%%%%%%%%%%%%%%%%%%%%%%%%%%%%%%
\section{Introduction}
\label{sec:intro}

Future gravitational wave (GW) observatories --
such as
the Laser Interferometer Space Antenna (LISA)~\cite{2017arXiv170200786A,LISACosmologyWorkingGroup:2022jok} or
the Einstein Telescope (ET)~\cite{Punturo:2010zz} --
could probe yet unexplored epochs of the cosmic evolution.
In particular, detecting
a stochastic gravitational wave background (SGWB)~\cite{Caprini:2018mtu} sourced in
the early Universe would provide evidence for new physics.
Such a background of gravitational radiation could e.g.~originate from
a cosmological first-order phase transition (PT)~\cite{Witten:1984rs,Hogan:1983ixn,PhysRevLett.65.3080,PhysRevD.49.2837}, as predicted by many theories beyond the Standard Model (SM).
Examples of such theories are scalar SM extensions~\cite{
  Profumo:2007wc,Ashoorioon:2009nf,Espinosa:2011ax,Profumo:2014opa,Balazs:2016tbi,
  Kurup:2017dzf,Chen:2017qcz,Chiang:2018gsn,Carena:2019une,Niemi:2021qvp,Azatov:2022tii} or
secluded dark sectors~\cite{
  Schwaller:2015tja,Breitbach:2018ddu,Helmboldt:2019pan,Bigazzi:2020avc,
  Reichert:2021cvs,Ertas:2021xeh,Morgante:2022zvc}.

Particularly motivated SM extensions are classically conformal (CC) theories
which are the focus of this work.
Their key feature is the absence of dimensionful terms in the Lagrangian.
As a consequence, the breaking of the electroweak (EW) symmetry is realized dynamically, either by
radiative corrections~\cite{
  Meissner:2006zh,Foot:2007iy,Espinosa:2008kw,Iso:2009ss,Iso:2009nw,Iso:2012jn,
  Farzinnia:2013pga,Englert:2013gz,Hashimoto:2013hta,Khoze:2014xha} or
strong dynamics~\cite{
  Hur:2011sv,Heikinheimo:2013fta,Holthausen:2013ota,Kubo:2014ova,Ametani:2015jla,
  Kubo:2015joa,Hatanaka:2016rek,Baratella:2018pxi}.
Classically conformal theories therefore not only alleviate the hierarchy problem \cite{Bardeen:295811},
but can also
account for dark matter~\cite{
  Hambye:2013dgv,Carone:2013wla,Khoze:2013uia,Steele:2013fka,Benic:2014aga,Guo:2015lxa,Oda:2017kwl,Hambye:2018qjv,Baldes:2021aph,Kawana:2022fum,Khoze:2022nyt,Frandsen:2022klh},
generate the baryon asymmetry of the Universe~\cite{
  Konstandin:2011dr,Konstandin:2011ds,Servant:2014bla,Khoze:2013oga,Baldes:2021vyz,Huang:2022vkf,Dasgupta:2022isg}, and
produce a SGWB~\cite{
  Jinno:2016knw,Kubo:2016kpb,Marzola:2017jzl,Iso:2017uuu,
  Marzo:2018nov,Aoki:2019mlt,Prokopec:2018tnq,Ellis:2020nnr,Wang:2020jrd,Kierkla:2022odc}.

In classically conformal models, the Universe may undergo
a vastly different cosmological history than predicted by the SM \cite{Kajantie:1996mn}.
A period of thermal inflation can be induced, while the Higgs field remains trapped in the unbroken phase.
If this epoch lasts until the temperature of the thermal bath approaches the scale of quantum chromodynamics (QCD),
there can be a first-order chiral phase transition ($\chi$PT) with six massless flavors~\cite{Pisarski:1983ms,Brown:1990ev} which subsequently triggers
electroweak symmetry breaking (EWSB)~\cite{
  Witten:1980ez,Iso:2017uuu,vonHarling:2017yew,Bodeker:2021mcj,Arunasalam:2017ajm,Ipek:2018lhm,Croon:2019ugf}.

In this work, we study the scenario where the combined QCD-EW phase transition initiates
the exit from supercooling.
Then a large amount of energy is naturally released during a supercooled transition.
The duration of the PT, on the other hand, is governed by the nucleation rate of hadronic bubbles.
Hence, it is crucial to model the strong dynamics.
To this end, we resort to different
low-energy quark-based QCD effective models which feature a first-order transition.
Such models have mainly been employed in the context of
strongly coupled hidden sectors~\cite{Helmboldt:2019pan,Reichert:2021cvs,Croon:2019iuh}.
Because of the short transition timescale, the associated GW signatures are typically weak.
Conversely, we show that our setup enhances the duration of the PT
which significantly improves observational prospects.
Our analysis remains largely model independent regarding the field content of the extended SM, which
renders our results applicable to a wide landscape of classically conformal theories.

Our work is structured as follows.
Section~\ref{sec:model} motivates our setup with a discussion of the dynamics which arise
in scale-invariant extensions of the SM.
Subsequently, we discuss different low-energy quark-based QCD effective models
in Sec.~\ref{sec:effective_QCD}.
This includes the Nambu--Jona-Lasinio (NJL) model and
two Polyakov loop extended versions thereof.
With these tools at hand, we compute the dynamics of the chiral phase transition in
a supercooled Universe
in Sec.~\ref{sec:chiralPT}.
Finally, Sec.~\ref{sec:GW_signal} estimates
the expected GW signature and discusses their observational prospects.
Lastly, some technical details of effective QCD theories are collected
in Sec.~\ref{sec:WF_renormalisation}, while we discuss the robustness and model dependence of our results
in Sec.~\ref{sec:cutoff_scheme_dependence}. In addition, we provide access
to our results in the Supplemental Material.

%%%%%%%%%%%%%%%%%%%%%%%%%%%%%%%%%%%%%%%%
%%%%%%%%%%%%%%%%%%%%%%%%%%%%%%%%%%%%%%%%
\section{Dynamics of classically conformal SM extensions}
\label{sec:model}

Classically conformal SM extensions
exhibit interesting dynamics at or above the electroweak scale.
This section reviews their most important aspects in a model-independent fashion.
The key feature of this class of models is an
additional gauge symmetry besides the SM gauge groups while
imposing scale invariance at tree level.
Consequently, quadratic terms are absent in the Lagrangian and the scalar sector of a CC extended SM reads
\begin{align}
\label{eq:V_tree}
    V(\Phi,H) =
        \lh (H^\dagger H)^2 &
      + \lphi (\Phi^\dagger \Phi)^2
    \nn
    &- \lp (\Phi^\dagger \Phi) (H^\dagger H)
    \;,
\end{align}
where
$H$ is the SM Higgs doublet and
$\Phi$ a beyond the Standard Model (BSM) scalar field.
Hence, the SM Higgs mass term is replaced by
a portal coupling, $\lp$, to
the new field $\Phi$
\begin{equation}
    -\mu_H^2 H^2 \to -\lp \Phi^2 H^2
    \;.
\end{equation}
\noindent
The characteristic feature is that EWSB is now tied to the dynamics of the conformal sector.
In contrast to the SM where EWSB is realized by a negative mass term,
the SM Higgs potential is recovered via the negative portal coupling once $\Phi$ acquires
a vacuum expectation value (VEV) $\vphi$.
Then we have $\mu_H^2 \simeq \lp \vphi^2$ and the EW symmetry is broken
in analogy to the SM, with the difference that the EW scale is now generated dynamically.
In the following, we assume that the conformal symmetry breaking occurs radiatively via
the Coleman-Weinberg mechanism~\cite{Coleman:1973jx}.
The same outcome could also be achieved by strongly coupled new physics \cite{vonHarling:2017yew}.

The scale of radiatively broken theories is characterized by
the gauge boson mass $M$ after symmetry breaking, as one typically has
$M \sim g \vphi$ with
$g$ the corresponding gauge coupling.
The lack of evidence for new particles below the electroweak scale hints at
$\vphi \gg v_H = 246~\GeV$.
Therefore, the symmetry breaking in a classically conformal setup is expected to occur first along
the $\Phi$ direction, with a critical temperature
$T_{{\rm c},\Phi} > T_\rmii{EW}$.
Then the first PT is well described by considering the BSM sector independently~\cite{Prokopec:2018tnq}.

Transitions in a multifield space exhibit a few typical outcomes.
At large temperatures, both scalars sit at the origin, and the potential
in the $\Phi$ direction is
\begin{equation}
    \Veff(\Phi,T) =
        \lphi (\Phi^\dagger \Phi)^2
      + V_\rmii{CW}(\Phi)
      + V_\T(\Phi,T)
    \;,
\end{equation}
where $V_\rmii{CW}(\Phi)$ is the temperature-independent one-loop contribution that induces a VEV for $\Phi$.
The finite-temperature corrections are contained in $V_\T(\Phi,T)$, which renders a thermal barrier separating the true from the metastable vacuum.
This barrier persists down to $T \to 0$ due to the conformal invariance at tree level.
If the tunneling rate is sufficiently small,
classically conformal SM extensions can therefore feature a large amount of supercooling. Then, the system remains trapped in the false vacuum at temperatures well below the critical temperature. As a consequence, the EW symmetry remains unbroken at small temperatures.

Initially, the Universe is radiation dominated at high temperatures. During supercooling, the false vacuum energy of the extended SM eventually becomes sizable compared to the thermal bath, and starts to dictate the cosmic expansion rate.
We expect an era of thermal inflation%
\footnote{
  Thermal inflation typically lasts only a small number of
  $e$-folds and is not to be mistaken with the period of cosmic inflation;
  see e.g.~\cite{Lyth:1998xn} for a review.
} starting at the temperature $\Ti$ defined by
\begin{equation}
\label{eq:Ti_definition}
    \Veff(0,\Ti) = \frac{\pi^2}{30} g_{\star,\epsilon} \Ti^4
    \;,
\end{equation}
where $\Veff$ is shifted such that
$\Veff(\vphi,T=0) = 0$.
The energetic degrees of freedom
$g_{\star,\epsilon} > 106.75$ depend on the particle content of the extended SM.
The thermal inflationary epoch and therefore supercooling terminates when $\Phi$
tunnels to the true vacuum at the temperature $\Tp < \Ti$.
The false vacuum energy is transferred back to the radiation bath and
the plasma is reheated to a temperature $\TRH$ close or equal to
the inflationary temperature
$\TRH \lesssim \Ti$. $\TRH$ is determined by the duration of the reheating period.

Depending on the scale of $\Tp$, different thermal histories can be realized:
\begin{itemize}
  \item
    $\Tp > T_\rmii{EW,SM} > \TQCD$\\[1.5mm]
    The conformal PT occurs above the electroweak scale,
    while the EW symmetry remains unbroken due to temperature corrections.
    Therefore, all dynamics below $\Tp$, such as the EW and chiral symmetry breaking ($\csb$), proceed
    as in the SM.
  \item
    $T_\rmii{EW,SM} > \Tp > \TQCD$\\[1.5mm]
    EWSB is delayed to a temperature $\Tp$ well below $T_\rmii{EW,SM}$.
    The conformal and EW symmetry breaking occur simultaneously.
    Subsequently, the EW symmetry may be restored if
    $\TRH > T_\rmii{EW,SM}$.
    The QCD scale is untouched by the new physics.
  \item
    $T_\rmii{EW,SM} > \TQCD \geq \Tp$\\[1.5mm]
    The system remains trapped in the unbroken phase all the way to the QCD scale.
    First, $\csb$ occurs, and triggers the EW and eventually the conformal PT.
\end{itemize}
We are interested in the last scenario. In fact, this outcome is expected in a large fraction of the parameter space,
e.g.\ for
$\gbl \lesssim 0.25$ in
the $\UBL$ extended SM~\cite{Marzo:2018nov,Ellis:2020nnr}.
Including QCD effects then induces peculiar dynamics which we now demonstrate.

Since the EW symmetry remains unbroken, $\csb$ occurs with six massless flavors.
Quark condensation then takes place at $\TQCD \sim 85~\MeV$~\cite{Braun:2006jd}.
In the massless limit,
different symmetry breaking patterns can be realized~\cite{Bodeker:2021mcj},
depending on if the transition is rendered first or second order.
Below, we assume a first-order $\chi$PT~\cite{Pisarski:1983ms,Brown:1990ev}.
When quark condensation takes place, we have
\begin{equation}
    q \bar{q} \to \langle q \bar{q}\rangle \neq 0
    \; .
\end{equation}
Because of the bilinear coupling of the quarks to the SM Higgs field $\sim y_q q\bar{q} H$,
the Higgs potential is destabilized and
\begin{equation}
    V(\Phi=0,h,T \leq \TQCD) \simeq
        \frac{1}{4} \lh h^4
      + \frac{y_t}{\sqrt{2}} \langle t\bar{t} \rangle  h \; ,
\end{equation}
where only the top quark contribution is considered.
To remain model independent, we neglect corrections to the Higgs potential stemming from both
vacuum loops as well as
thermal effects.
Therefore, QCD necessarily triggers the breaking of the electroweak symmetry~\cite{Iso:2017uuu,vonHarling:2017yew,Bodeker:2021mcj} as
the Higgs field acquires a VEV,
\begin{equation}
    \vhqcd = \biggl[-\frac{y_t}{\sqrt{2}\lambda_H} \langle t\bar{t}\rangle \biggr]^{1/3}
    \; .
\end{equation}
In turn, the Higgs VEV backreacts onto the BSM direction via
the $\lp$-portal term in Eq.~\eqref{eq:V_tree} and
\begin{equation}
\label{eq:Veff_after_Tqcd}
    V^{ }_{T\leq T_\rmii{QCD}}(\Phi,T) =
    V^{ }_{T > T_\rmii{QCD}}(\Phi,T)
      - \frac{\lp}{2} \vhqcd^2 \Phi^2
        \; .
\end{equation}
Since the sign of the portal coupling is negative, the additional term counteracts the thermal barrier. This leads to one of the following scenarios, schematically displayed in Fig.~\ref{fig:overview}:
\begin{figure*}[t]
    \centering
    \includegraphics[width=0.9\textwidth]{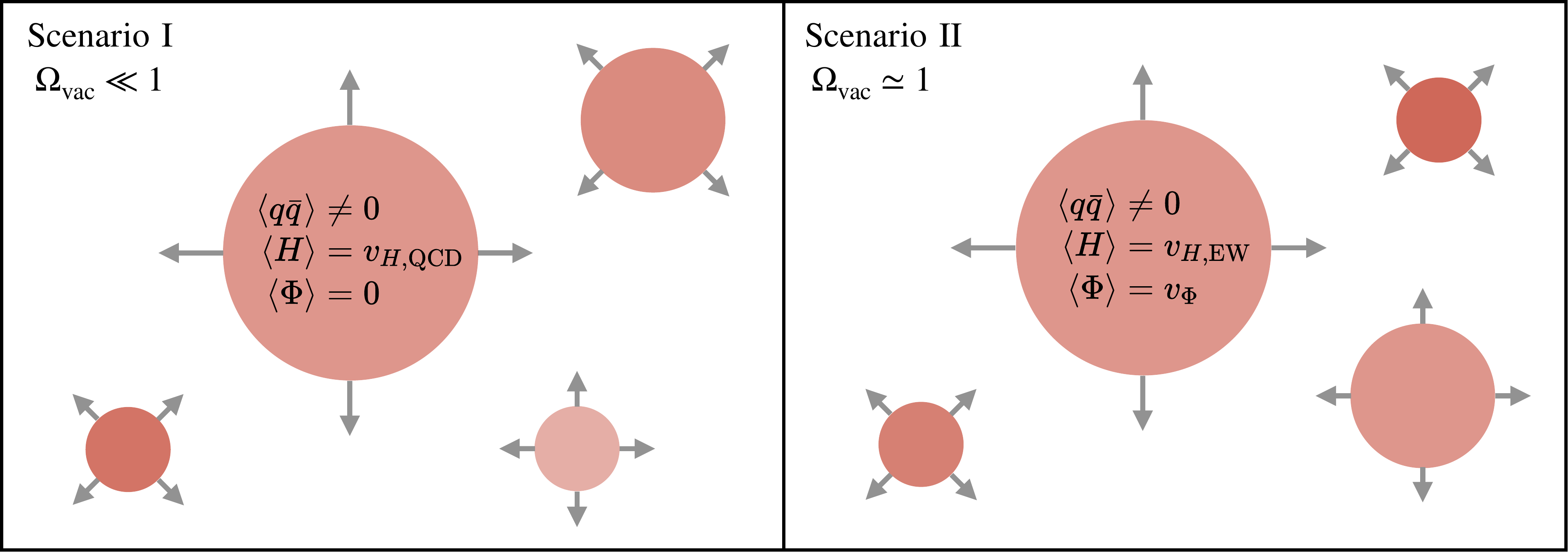}
    \caption{%
    Schematic overview of
    scenarios \ref{scenarioI} and \ref{scenarioII}
    at $T = \TQCD$.
    Scenario~\ref{scenarioI}:
    $\Phi$ remains in the metastable vacuum after $\csb$. Therefore, the Universe keeps inflating until the conformal sector undergoes a PT at a temperature well below $T_\mathrm{QCD}$.
    The energy budget of $\csb$ is solely sourced by QCD and hence small
    $\Omega_\mathrm{vac} \simeq \rho_\mathrm{vac,\mathrm{QCD}}/(\rho_\mathrm{vac,BSM} + \rho_\mathrm{rad}) \ll 1$.
    Scenario~\ref{scenarioII}:
    quark condensation triggers the PT in the conformal sector.
    The false vacuum energy responsible for the thermal inflation is released in the transition, thus
    $\Omega_\mathrm{vac} \simeq \rho_{\mathrm{vac,BSM}}/(\rho_\mathrm{vac,BSM}+\rho_\mathrm{rad}) \simeq 1$.
    }
    \label{fig:overview}
\end{figure*}
\begin{enumerate}[{\bf I.}]
  \item
    \label{scenarioI}
    The height of the barrier in $\Veff (\Phi,T)$ shrinks and the onset of the conformal transition is accelerated. Then $\Phi$ tunnels at a temperature close to, but well below, $\TQCD$.
  \item
    \label{scenarioII}
    The QCD term in Eq.~\eqref{eq:Veff_after_Tqcd} cancels the barrier in $\Veff (\Phi,T)$, thus triggering the breaking of the conformal symmetry and initiates the end of the supercooling epoch right at $\TQCD$.
    Subsequently, the Higgs field takes its SM VEV $v_H = 246~\GeV$ as $\Phi$ rolls down to $\vphi$.
\end{enumerate}

In the following, we explore scenario~\ref{scenarioII}.
We assume that quark condensation first occurs in the light quark sector~\cite{Bodeker:2021mcj} in a first-order transition. In this case, we further assume that the impact of the Higgs field on the $\chi$PT is negligible. Eventually, however, the Higgs potential is destabilized as the top condensate forms. As a consequence, $\Phi$ rolls down to its potential minimum inside the hadronic bubbles. Then, also the Higgs evolves towards its SM VEV.

As the QCD transition marks the exit from supercooling, the PT involves a large amount of latent heat which may generate a sizable stochastic gravitational wave background.
This energy budget is determined by the false vacuum energy of the scale-invariant sector which drives supercooling and can thus be expressed via $\Ti$ (cf.\ Eq.~\eqref{eq:Ti_definition}) in a model-independent fashion.
The false vacuum energy can, however, only be released in patches of space where the chiral symmetry is broken. Therefore, the initial stage of the PT, namely the nucleation of bubbles, is governed by QCD.
Hence, it is crucial
to accurately model the strong dynamics to understand the transition.

%%%%%%%%%%%%%%%%%%%%%%%%%%%%%%%%%%%%%%%%
%%%%%%%%%%%%%%%%%%%%%%%%%%%%%%%%%%%%%%%%
\section{Effective description of QCD}
\label{sec:effective_QCD}

We introduce
two quark-based effective QCD models
with $\Nf = \Nc = 3$.
Both models feature a
$\SU(3)_\rmii{L} \times \SU(3)_\rmii{R}$ chiral symmetry, which is broken down to
$\SU(3)_\rmii{V}$ during $\csb$. The choice of installing $\Nf = 3$ reflects our assumption that the quarks condense first in the light sector.
While the proposed models with $\Nf = 3$ do not reflect the symmetry of full QCD,
they are also convenient since then the model parameters can be fixed by hadronic physics.

The main findings of our work are largely independent of
the concrete EFT description of QCD,
as we will stress later.
The only requirement for the applicability of our mechanism is that
the model exhibits a first-order transition.
For a recent review on QCD phase transitions see~\cite{Aarts:2023vsf}.

\subsection{Nambu--Jona-Lasinio (NJL) model}

To study the strong dynamics in our setup, we employ
a Nambu--Jona-Lasinio (NJL) model~\cite{Nambu:1961fr} with $\Nf = 3$ flavors;
see e.g.~\cite{Klevansky:1992qe,Buballa:2003qv} for reviews.
The free parameters of the model are fitted to recover
the correct meson masses and decay constants of QCD.
Then, we take the chiral limit to approximate
scenario~\ref{scenarioII} introduced in Sec.~\ref{sec:model}.

The Lagrangian density of
the NJL model reads~\cite{Kunihiro:1989my,Rehberg:1995kh}
\begin{align}
\label{eq:NJL:lag}
    \mathcal{L}_\rmii{NJL} &=
      \bar{q} (i\msl{\partial} - \hat{m}) q
    + \mathcal{L}_\rmii{4F}
    + \mathcal{L}_\rmii{6F}
    \;,
\end{align}
and
includes as multifermion interactions
$\mathcal{L}_\rmii{4F}$ and
the six-fermion determinant interaction
$\mathcal{L}_\rmii{6F}$ for three flavors:
\begin{align}
    \mathcal{L}_\rmii{4F} &=
    G \sum_{a=0}^{\Nf^2 -1} \Bigl[
        \left(\bar{q}\, T^a q\right)^2
      + \left(\bar{q}\, i\gamma_5 T^a q\right)^2
    \Bigr]
    \;,
    \\
    \mathcal{L}_\rmii{6F} &=
    G_\rmii{D} \Bigl[
        \mbox{det}\left(\bar{q}(1-\gamma_5) q \right)
      + \mbox{det}\left(\bar{q}(1+\gamma_5) q \right)
      \Bigr]
   \;,
\end{align}
with effective coupling parameters
$G$ and
$G_\rmii{D}$.
Here,
$T^a$ denote the generators of $\SU(\Nf)$, hence the sum in the second term runs from
$a = 0, 1,\dots,\Nf^2 - 1$.
The sextic interaction $\mathcal{L}_\rmii{6F}$ is known as
the 't~Hooft determinant~\cite{tHooft:1976rip,Kobayashi:1970ji,Kobayashi:1971qz}.
This determinant mimics the anomalous breaking of
the axial ${\rm U}(1)_\rmii{A}$ symmetry in QCD by including a six-point quark interaction which breaks
${\rm U}(1)_\rmii{A}$ explicitly.
This leaves the Lagrangian with an unbroken
$\SU(3)_\rmii{L} \times \SU(3)_\rmii{R} \times {\rm U}(1)_\rmii{V}$ symmetry.

The quark mass term $\sim \hat{m} \bar{q} q$ in Eq.~\eqref{eq:NJL:lag} breaks chiral symmetry explicitly.
With $q = (u,d,s)^T$,
the mass matrix reads
\begin{equation}
    \hat{m} =
    \mathrm{diag}(m_u,m_d,m_s) =
    \mathrm{diag}(y_u,y_d,y_s) \frac{\langle h \rangle}{\sqrt{2}}
    \;,
\end{equation}
where
$h$ is the physical SM Higgs.
As long as the electroweak symmetry remains unbroken, we have $\langle h \rangle = 0$.
Thus, the mass term in the Lagrangian vanishes and we obtain a
$\SU(3)_\rmii{L} \times \SU(3)_\rmii{R}$ symmetric expression.
This limit is referred to as the chiral limit, which we will consider once
all model parameters are fixed in the massive limit.

To perform computations within the NJL model,
typically the self-consistent mean-field approximation
(MFA)~\cite{Kunihiro:1983ej,Hatsuda:1994pi,Hatsuda:1985eb} is applied.
In its compact form, the NJL Lagrangian~\eqref{eq:NJL:lag}
is given by~\cite{Holthausen:2013ota}
\begin{align}
\label{eq:NJL:lag:mfa}
    \mathcal{L}_\rmii{NJL} &=
      \bar{q} \bigl(i\msl{\partial} - \hat{m}\bigr) q
    + \mathcal{L}_\rmii{4F}
    + \mathcal{L}_\rmii{6F}
    \;,\nn[1mm]
    \mathcal{L}_\rmii{4F} &=
      2 G\, \tr\,\bigl(\Psi^\dagger \Psi\bigr)
    \;,\nn[1mm]
    \mathcal{L}_\rmii{6F} &=
      G_\rmii{D} \bigl(\det \Psi + \mathrm{h.c.}\bigr)
    \;,
\end{align}
with the fermion bilinear $\Psi_{ij} = \bar{q}_j(1-\gamma_5) q_i$.
In the MFA, the Lagrangian is expressed by $\Psi$ and
the expectation value of the quark bilinear $\langle \Psi \rangle$,
which, in turn, can be expressed in terms of auxiliary meson fields
\begin{equation}
    -4 G \langle \Psi \rangle =
      \left(\sigma + i\eta'\right) \mathbbm{1}
    + 2\left(a_a + i\pi_a\right) T^a
    \;,
\end{equation}
where
$T^a$ are the $\SU(3)$ generators in the fundamental representation, and
\begin{equation}
\label{eq:Psi:mfa}
  \sigma = -\frac{4 G}{3} \langle \bar{q} q \rangle
    \;,
\end{equation}
for the scalar singlet
as well as
$\eta' \sim \langle \bar q \gamma_5 q \rangle$ for the pseudoscalar singlet,
$a_a \sim \langle \bar q\,T^a q \rangle$ the scalar octet, and
$\pi_a \sim \langle \bar q \gamma_5 T^a q \rangle$ the pseudoscalar octet,
which are related to the respective chiral condensates.%
\footnote{
  In general, these condensates are spatially inhomogeneous~\cite{Heinz:2015lua},
  e.g.\ $\sigma = \sigma(\vec{x})$,
  which is relevant at finite chemical potential.
  }
Consequently, the Lagrangian~\eqref{eq:NJL:lag:mfa} is split into
a mean-field term that contains only terms up to quadratic order in the fermion fields and
a term that encodes higher-order interactions;
the explicit NJL Lagrangian in the MFA is given in~\cite{Helmboldt:2019pan}.

Expressing the NJL Lagrangian in the MFA
via Eq.~\eqref{eq:Psi:mfa}, yields
the tree-level potential~\cite{Helmboldt:2019pan}
\begin{align}
  \label{eq:V0:NJL}
    V_0^\rmii{NJL} &(\sigma, \eta^\prime, a_a, \pi_a) =
    \\
    &+\frac{1}{8G}\left(
       3\sigma^2
     + 3\eta^{\prime 2}
     + 2\pi_a\pi_a
     + 2 a_a a_a\right)
    \nn
    &-\frac{G_\rmii{D}}{16 G^3} \left[\sigma(\sigma^2
      + \pi_a \pi_a
      - 3\eta^{\prime 2}
      - a_a a_a)
    + 5 a_a \pi_a \eta^\prime\right]
    \;.
    \nonumber
\end{align}
Since during $\csb$, only $\sigma$ can acquire a VEV,%
\footnote{
  Mesons that couple fermions of different flavor cannot acquire a VEV.
  Defining the effective scalar meson fields via their constituent fermions and given that
  the isospin symmetry ${\rm SU}(3)_\rmii{V}$ remains intact,
  only $\sigma$ can obtain a finite VEV since it is the only meson proportional to the identity matrix.
  Pseudoscalars cannot acquire a non-zero VEV because the vacuum is parity even.}
we have vanishing $\eta' = a_a = \pi_a = 0$, {\em viz.},
\begin{equation}
\label{eq:V0_sigma:NJL}
  V_0^\rmii{NJL} (\sigma) =
    \frac{3}{8 G} \sigma^2 - \frac{G_\rmii{D}}{16 G^3} \sigma^3
  \;.
\end{equation}
This is the tree-level Lagrangian relevant for studying
the chiral phase transition.

Radiative corrections at one-loop level are obtained
by integrating out the fermions.
This corresponds to computing
the fermion vacuum energy
\begin{equation}
\label{eq:NJL_oneloop}
    V_1^\rmii{NJL}(\sigma) =
      -\Nc \sum_i \int \frac{{\rm d}^4p}{i(2\pi)^4}
      \ln \det\left(\msl{p}-M_i\right)
    \;.
\end{equation}
In the chiral limit, only one of
the respective auxiliary meson field masses $M_i$ remains.
This corresponds to the effective quark mass~\cite{Fukushima:2008wg}
\begin{equation}
\label{eq:M:sigma}
  M_i \equiv M(\sigma) = \sigma - \frac{G_\rmii{D}}{8 G^2} \sigma^2
  \;.
\end{equation}
In principle,
the above expression contains
a linear-in-$H$ term $\sim y_i H$ with $H$ the SM Higgs doublet,
which vanishes in the chiral limit of our setup.

Because of the multifermion interactions in Eq.~\eqref{eq:NJL:lag},
the NJL model is nonrenormalizable.
Therefore, a regularization procedure is necessary by
truncating the model to the six-fermion operator and
installing a hard momentum cutoff $\Lambda$ as a model parameter.
Different regularization schemes are compared in~\cite{Kohyama:2015hix}.
While UV divergences,
such as the vacuum energy in Eq.~\eqref{eq:NJL_oneloop},
are rendered finite,
now the model crucially depends on the employed cutoff scheme.
A four-dimensional cutoff scheme is used in~\cite{Helmboldt:2019pan,Ametani:2015jla,Holthausen:2013ota,Aoki:2019mlt}
and inspected in our Appendix~\ref{sec:cutoff_scheme_dependence}, whereas
a three-dimensional scheme is used in~\cite{Hansen:2006ee,Reichert:2021cvs,Fukushima:2013rx}.
Because of the thermal aspect of the computation,
we employ a sharp 3D momentum cutoff in the main part of this work,
replacing all
\begin{equation}
    \int\frac{{\rm d}^4 p}{(2\pi)^4} \to
    \int\frac{{\rm d} p_0}{2\pi}\int^\Lambda_{\vec{p}} \; ,
\end{equation}
where the $d$-dimensional measure is
$\int_{\vec{p}} \equiv \mu^{3-d} \int \!
\frac{{\rm d}^d\vec{p}}{(2\pi)^d}$
with $d=3$. Such a sharp cutoff follows~\cite{Hansen:2006ee}
in contrast to a cutoff of only the fermion vacuum energy~\eqref{eq:NJL_oneloop}
\`a la~\cite{Fukushima:2013rx}.
To estimate the robustness of our results with regard to the cutoff scheme,
we compare to the 4D approach
in Appendix~\ref{sec:cutoff_scheme_dependence}.
With a 3D cutoff,
the one-loop part of the effective potential~\eqref{eq:NJL_oneloop} evaluates to
\begin{align}
\label{eq:NJL:1l:3d}
  \VNJL_{1,\rmii{3D}} (\sigma) &=
  -2\Nc\Nf\int_{\vec{p}}^{\Lambda} E_p
  \nn &=
  -\frac{\Nc \Nf }{8 \pi^2}\Lambda^{4}\biggl[
      \left(2 + \xi^2 \right)\sqrt{1 + \xi^2}
    \nn&
    \hphantom{{}=-\frac{\Nc \Nf }{8 \pi^2}\Lambda^{4}\biggl[}
    + \frac{\xi^{4}}{2} \ln \frac{\sqrt{1 + \xi^2} - 1}{\sqrt{1 + \xi^2} + 1}
    \biggr]
    \;.
\end{align}
with
$E_p = \sqrt{p^2 + M^2}$
and
$\xi = M/\Lambda$.
As a consequence,
the NJL model has three parameters $G$, $G_\rmii{D}$, and $\Lambda$.
By reproducing the hadron spectrum of QCD,
all three parameters can be fixed.
In Table~\ref{tab:NJL_params_QCD},
we summarize the employed values, taken from Ref.~\cite{Fukushima:2008wg}.
\begin{table}
\setlength\tabcolsep{6pt}
\begin{tabular}{ |c|c|c| }
\hline
$\Lambda\left[{\rm MeV}\right]$
& $G \Lambda^2$
& $G_\rmii{D} \Lambda^5$ \\
\hline
$631.4$ & $1.835$ & $-9.29$
\\
\hline
\end{tabular}\\\vspace{.2cm}
\begin{tabular}{|c|c|c|c|c|}
    \hline
    $a_0$ & $a_1$ & $a_2$ & $a_3$ & $T_\glue\;[{\rm MeV}]$ \\
    \hline
    $3.51$& $-2.47$ & $15.2$ & $-1.75 $& $178$ \\
    \hline
\end{tabular}
  \caption{%
    Upper panel:
    NJL parameters which reproduce the QCD meson spectrum~\cite{Fukushima:2008wg}. Here, $G$ has been rescaled via
    $G = g_\rmii{S}/2$, with
    $g_\rmii{S}$ from~\cite{Fukushima:2008wg}, to match our different coupling conventions.
    Lower panel:
    parameters for the pure glue part~\cite{Roessner:2006xn}.
    }
\label{tab:NJL_params_QCD}
\end{table}

\begin{figure}
    \centering
    \includegraphics[width=\columnwidth]{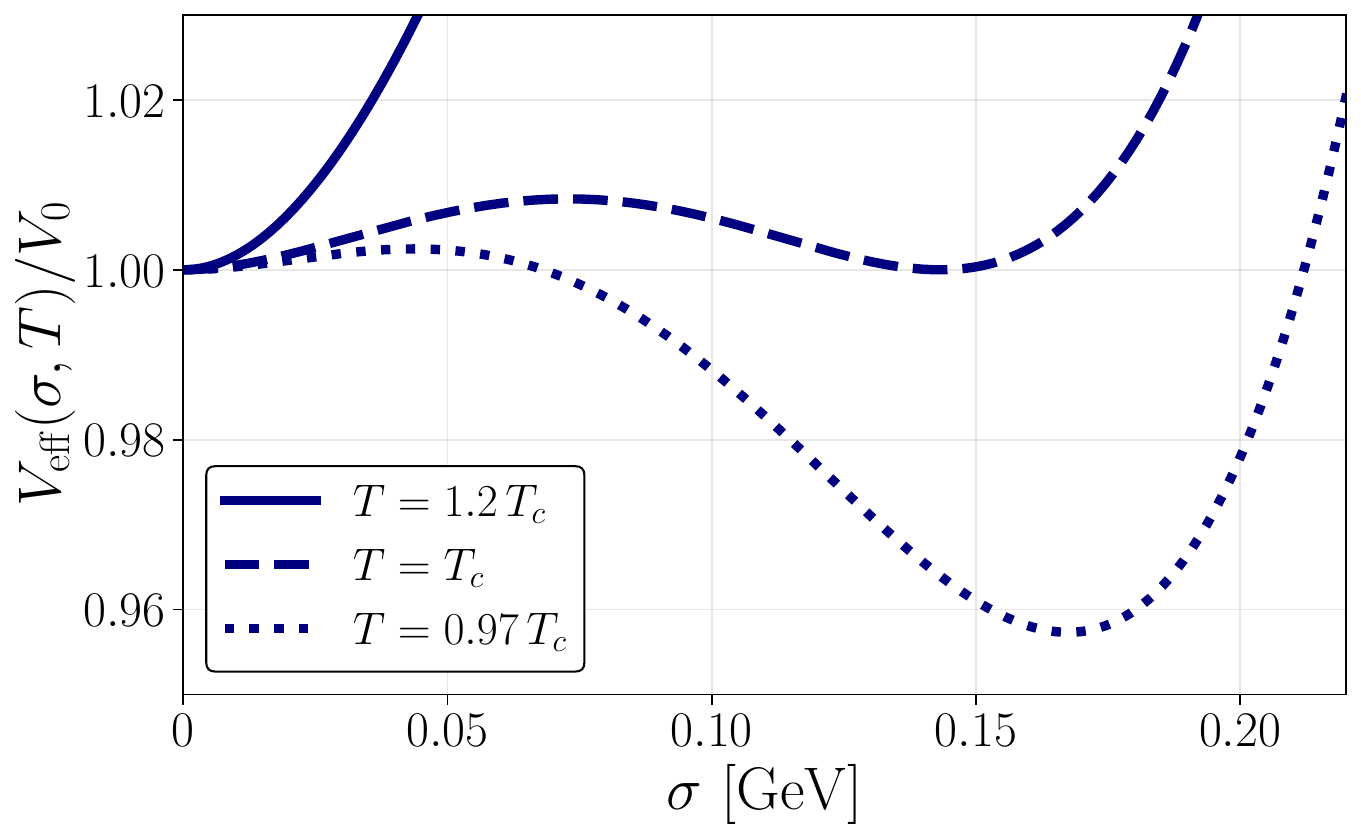}
    \caption{%
      Effective potential for the NJL model as a function of
      the chiral condensate $\sigma$ and
      temperature $T$, normalised to the false vacuum energy $V_0 = \Veff(\sigma = 0, T=0)$. The critical temperature is denoted by $\Tc$.
      At large~$T > \Tc$, the chiral symmetry is restored.
      As the Universe cools, a second minimum forms, which is separated from the origin by a thermal barrier, indicating a first-order transition. Both the PNJL model and its improved version show a similar symmetry breaking pattern.}    
    \label{fig:PNJL:Veff}
\end{figure}

At one-loop level and finite temperature,
the effective potential receives another contribution from the interaction with
the thermal medium.
Since the degrees of freedom are fermionic,
we can incorporate this contribution via
\begin{equation}
\label{eq:NJL:1lT}
  \VNJL_\T(\sigma,T) = g_q \Nc J_\rmii{F} (M^2/T^2)
  \;,
\end{equation}
where
the number of quark degrees of freedom is
$g_q =
\Nf~({\rm quarks}) \times
2~({\rm antiquarks}) \times
2~({\rm spin}) = 12$
and
the fermionic thermal integral is
\begin{align}
\label{eq:J:BF}
  J_\rmii{F} (M,T) &=
    - T\int_{\vec{p}}\ln\Bigl[1 + e^{-E_p/T}\Bigr]
    \;.
\end{align}
This completes the construction of the effective potential for the NJL model.
To summarize, we have
\begin{equation}
    \VeffNJL (\sigma,T) =
        \VNJL_0 (\sigma)
      + \VNJL_1(\sigma)
      + \VNJL_\T(\sigma,T)
      \;.
\end{equation}
The potentials for the employed quark-based QCD effective models,
are visualized in Fig.~\ref{fig:PNJL:Veff} for different values
of
the chiral condensate $\sigma$ and
temperature $T$.

\begin{figure}
    \centering
    \includegraphics[width=\columnwidth]{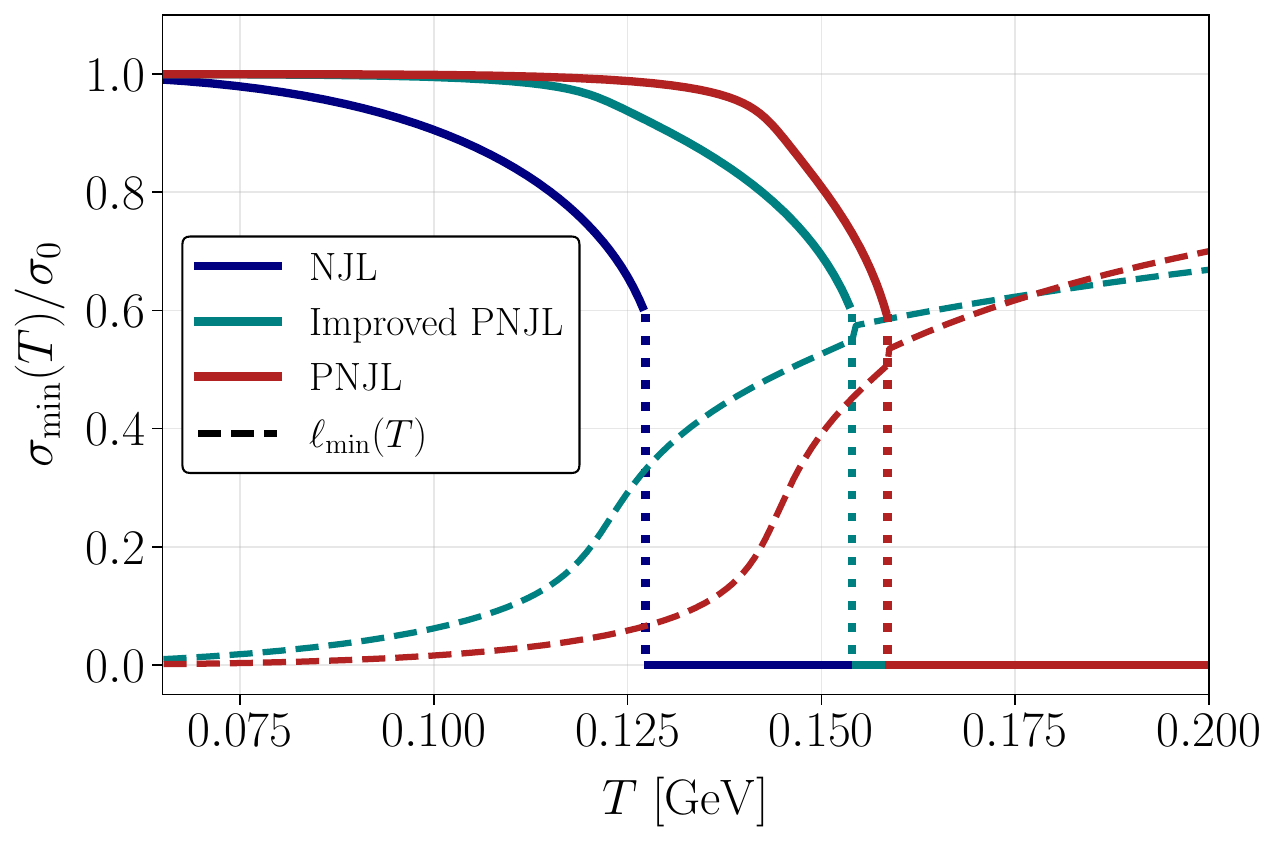}
    \caption{%
      Evolution of the chiral condensate $\sigma$ (solid) and
      the Polyakov loop $\ell$ (dashed) as a function of temperature
      for different quark-based QCD effective models.
      Here, $\sigma_0$ denotes the potential minimum at $T=0$.
      All models feature a first-order chiral transition, as
      the quark condensate jumps from zero to a finite value (dotted).
    }
    \label{fig:chiral_condensate_T}
\end{figure}

For the NJL model in Fig.~\ref{fig:chiral_condensate_T}, we display the evolution of
the chiral condensate $\sigma$ (blue) as a function of temperature.
The NJL model features a discontinuity at
$\Tc \approx 128~\MeV$, indicating a first-order transition.

\subsection{Polyakov loop enhanced NJL (PNJL) model}
Along the phase transition of $\csb$, the gluon dynamics is characterized by the order parameter $\ell$, which is the fundamental traced Polyakov loop \cite{Polyakov:1975rs}
\begin{align}
  \ell(\vec{x}) &= \frac{1}{\Nc} \tr_\rmi{c}\,\vec{L}
  \;,\\
  \vec{L}(\vec{x}) &= \mathcal{P} \exp\Bigl[ i\gs\!\int_{0}^{1/T}\!\!\! {\rm d}\tau A_4(\vec{x},\tau) \Bigr]
  \;,
\end{align}
where
$\tr_\rmi{c}$ is the trace in color space,
$\mathcal{P}$ denotes the path ordering along the time direction,
$\gs$ is the strong coupling,
$\bar\ell$ will be the charge conjugated Polyakov loop, and
$A_4 = iA_0$ is the Euclidean temporal component of the gauge field.
In the mean-field approximation,
we take $A_4$ spatially homogeneous~\cite{Meisinger:2001cq}.

The dynamics of the Polyakov loop can be included in the NJL model;
see e.g.~\cite{Fukushima:2017csk}
for an extensive review of the Polyakov loop.
Its effect is included by augmenting the NJL Lagrangian~\eqref{eq:NJL:lag}
by a pure gauge part~\cite{Meisinger:1995ih,Fukushima:2003fw,Ratti:2005jh}
\begin{equation}
\label{eq:PNJL:lag}
    \mathcal{L}_\rmii{PNJL} =
    \mathcal{L}_\rmii{NJL}
  - V_\rmi{glue} (\ell,\bar\ell,T)
  \;,
\end{equation}
where
$\partial_\mu \to D_\mu$ is now the covariant derivative and
$V_\rmi{glue}$
contributes as the Polyakov-loop potential
in the effective potential.

While the gluon potential
$V_\mathrm{glue} (\ell,\bar\ell,T)$
cannot be computed from first principles,
it can be parametrized and effectively fitted to lattice data.
Different parametrizations of the Polyakov loop potential are conceivable.
Below we employ one logarithmic parametrization that accounts for
the Haar measure~\cite{Roessner:2006xn,Fukushima:2017csk}
which reads
\begin{align}
  \label{eq:L:param}
    V_\mathrm{glue} (\ell,\bar{\ell},T) &=
      T^4 \Bigl[-\frac{1}{2} a(T) \ell\bar{\ell}
      \\ &
      + b(T)
      \ln\bigl(1
        - 6\ell\bar{\ell}
        + 4 \left(\ell^3 + \bar{\ell}^3\right)
        - 3 (\ell\bar{\ell})^2
      \bigr)
    \Big]
    \;.
    \nonumber
\end{align}
Here,
the expectation value of the Polyakov loop is restricted to lie within
the interval
$\ell = 0$ (confined) and
$\ell = 1$ (deconfined).
Because of this interval,
utilizing Eq.~\eqref{eq:L:param}
is more convenient for our numerical studies compared to
a polynomial parametrization~\cite{Ratti:2005jh,Fukushima:2017csk},
which was e.g.\ used in Refs.~\cite{Huang:2020crf,Reichert:2021cvs}.
However,
our results do not depend crucially on the employed parametrization.

The coefficients of Eq.~\eqref{eq:L:param}
are temperature dependent and parametrizable as
\begin{align}
  a(T) &= a_0
    + a_1 \left( \frac{T_\mathrm{glue}}{T} \right)^{ }
    + a_2 \left( \frac{T_\mathrm{glue}}{T} \right)^{2}
  \;, \\
  b(T) &= b_3 \left( \frac{T_\mathrm{glue}}{T} \right)^3
  \; ,
\end{align}
where $T_\rmi{glue} = 178~\MeV$~\cite{Schaefer:2007pw} is
the temperature associated with the breaking of
the $Z_3$ center symmetry of $\SU(3)$~\cite{McLerran:1981pb},
hence the confinement scale.
The parameters $a_0,\dots,a_2$ and $b_3$
are then fitted to lattice data. In Table~\ref{tab:NJL_params_QCD}, we collect the employed parameters.

The next-to-leading order (NLO) effective potential in the PNJL model can be expressed as
\begin{align}
  \label{eq:V:pnjl}
    \VeffPNJL (\sigma,T) &=
        \VPNJL_0 (\sigma)
      + \VPNJL_1(\sigma)
      \nn&
      + \VPNJL_\T(\sigma,\ell,\bar\ell,T)
      + V_\rmi{glue}(\ell,\bar\ell,T)
      \;,
\end{align}
where
\begin{align}
  \VPNJL_0 (\sigma) = \VNJL_0 (\sigma)
  \;,\quad
  \VPNJL_1 (\sigma) = \VNJL_1 (\sigma)
  \;.
\end{align}
Besides the novel pure gauge potential
$V_\rmi{glue}(\ell,\bar\ell,T)$,
in the PNJL model
also the one-loop medium interaction in Eq.~\eqref{eq:NJL:1lT}
is affected by the Polyakov loop.
A Polyakov loop in the fundamental representation
yields~\cite{Reichert:2021cvs}
\begin{align}
  \label{eq:PNJL:1l:T}
  \VPNJL_\T &(\sigma,\ell,\bar\ell,T) =
    \VNJL_\T(\sigma,T)
    \\ &
    - g_q T\int_{\vec{p}}
    \ln\Bigl[1
        + (3\ell - 1)e^{-E_p/T}
        + e^{-2 E_p/T}
      \Bigr]
    \;,
    \nonumber
\end{align}
which results from
including a suppression of the thermal weight with the Polyakov loop $\ell$.
Here,
$E_p = \sqrt{p^2 + M^2}$ and
$M = M(\sigma)$ from Eq.~\eqref{eq:M:sigma}.

We observe that the expectation value of the chiral condensate undergoes a jump at
$\Tc \approx 159~\MeV$ (cf.\ Fig.~\ref{fig:chiral_condensate_T}), while
the Polyakov loop evolves smoothly with temperature.

\subsection{Improved Polyakov loop potential}

One immediate shortcoming
of the PNJL model is that
the parameters of the Polyakov loop potential $V_\rmi{glue}$
are fitted against pure Yang-Mills data instead of full QCD.
Since the presence of quarks alters
the thermodynamic properties of the system,
their effect should be included.
This can be achieved by a rescaling of
the temperature~\cite{Haas:2013qwp}.

To this end,
we define the reduced temperatures
\begin{equation}
    t_\rmi{glue} = \frac{T - T^\rmi{glue}_\rmi{cr}}{T^\rmi{glue}_\rmi{cr}}
    \; , \quad
    t_\rmii{YM} = \frac{T - T^\rmii{YM}_\rmi{cr}}{T^\rmii{YM}_\rmi{cr}}
    \; ,
\end{equation}
with the absolute temperature scales
$T^\rmii{YM}_\rmi{cr} = 270~\MeV$ \cite{Fister:2013bh}
being the critical temperature in pure Yang-Mills, and
$T^\rmi{glue}_\rmi{cr}$
the critical temperature of a QCD-like theory~\cite{Schaefer:2007pw}.
In our case,
$T^\rmi{glue}_\rmi{cr} = 178~\MeV$ for
$\Nf = 3$ massless flavors.
With these definitions, the temperature-dependent terms in the Polyakov loop potential are rewritten as
\begin{equation}
    \frac{T_\rmi{glue}}{T} \to
    \frac{1}{1+t_\rmi{glue}}
    \;.
\end{equation}
For a given reduced glue temperature
$t_\rmi{glue}$ the corresponding Polyakov loop potential
$V_\rmi{glue}$ can be found by rescaling~\cite{Haas:2013qwp}
\begin{equation}
  t_\rmii{YM}(t_\rmi{glue}) \approx 0.57 \, t_\rmi{glue}
  \; .
\end{equation}
The
glue potential is then obtained by
\begin{equation}
    \frac{V_\rmi{glue}(\ell,\bar\ell,t_\rmi{glue})}{T_\rmi{glue}^4} =
    \frac{V_\rmii{YM}(\ell,\bar\ell,t_\rmii{YM}(t_\rmi{glue}))}{T_\rmii{YM}^4}
    \; .
\end{equation}
Together with this temperature rescaling,
another variant of the NJL is established.
In the following,
it is referred to as
the improved PNJL model.
As shown in Fig.~\ref{fig:chiral_condensate_T}, we find a slightly lower critical temperature
$\Tc \approx 154~\MeV$ compared to the conventional PNJL model.

%%%%%%%%%%%%%%%%%%%%%%%%%%%%%%%%%%%%%%%%
%%%%%%%%%%%%%%%%%%%%%%%%%%%%%%%%%%%%%%%%
\section{Supercooled chiral phase transition}
\label{sec:chiralPT}

When investigating the dynamics of the chiral phase transition for the different models,
we assume that the Universe undergoes an era of thermal inflation prior to $\csb$,
driven by a conformal sector; see Sec.~\ref{sec:model}.
For a model-independent analysis,
we characterize the new physics merely by the temperature $\Ti$ where
the false vacuum energy starts to dominate.
The Hubble parameter at the QCD scale is therefore approximately constant
and
\begin{align}
\label{eq:Hubble_Ti}
    H(\TQCD) &=
    \left(\frac{\rho_\mathrm{vac} + \rho_\mathrm{rad}(\TQCD)}{3 \MPl^2}\right)^\frac{1}{2}
    \nn
    &\simeq
    \left(\frac{\pi^2}{90} g_{\star,\epsilon}(\Ti) \frac{\Ti^4}{\MPl^2} \right)^\frac{1}{2}
    \simeq H(\Ti)
    \; ,
\end{align}
where the second line is valid for large
$\Ti \gg \TQCD$, and $\rho_\mathrm{vac} = \Veff$ from Eq.~\eqref{eq:Ti_definition}.
While the energetic degrees of freedom are set to the SM value
$g_{\star,\epsilon}(\Ti) = 106.75$,
the exact value is slightly larger as it depends on
the concrete particle content of the SM extension.
We then study scenario~\ref{scenarioII} from Sec.~\ref{sec:model},
where the chiral phase transition initiates the exit from supercooling.

The key quantity that sets
the PT dynamics is the false vacuum decay rate~\cite{Linde:1980tt,Callan:1977pt,Coleman:1977py,Hogan:1983ixn,Witten:1984rs}
of the chiral condensate, which is approximated by
\begin{equation}
\label{eq:false_vacuum_decay_rate}
    \Gamma(T) = T^4
      \biggl(\frac{S_3}{2\pi T}\biggr)^\frac{3}{2}
      \exp\biggl(-\frac{S_3}{T}\biggr)
      \; .
\end{equation}
Here,
$S_3$ denotes the three-dimensional Euclidean bounce action~\cite{Helmboldt:2019pan,Gould:2021ccf}
\begin{equation}
\label{eq:bounce_action}
    S_3 \left[\sigma\right] = 4\pi \int {\rm d}r\, r^2 \biggl[
        \frac{Z_\sigma^{-1}}{2} \left(\frac{{\rm d}\sigma}{{\rm d}r}\right)^2
      + \Veff(\sigma,T)
    \biggr]
  \;,
\end{equation}
where $\Veff(\sigma,T)$ is the (P)NJL effective potential from Sec.~\ref{sec:effective_QCD}.
Regarding the Polyakov loop extended models, we take $\ell = \bar\ell$ since we have zero chemical potential \cite{Fukushima:2017csk}. Then we determine $\ell$ such that
$\Veff(\sigma,\ell,T)$ is minimized for every combination of
$(\sigma, T)$, i.e.\
$\Veff(\sigma,T) = \Veff(\sigma,\ell_\mathrm{min},T)$.
In the effective theories considered here, $\sigma$ is not a fundamental degree of freedom
and classically nonpropagating.
Therefore, the kinetic term is generated at loop level and
the bounce action is augmented by the wave function renormalization
$Z_\sigma^{-1} = Z_\sigma^{-1}(\sigma,\ell,\bar\ell,T)$.
The computation of $Z_\sigma$ in the 3D cutoff scheme is detailed in
Appendix~\ref{sec:WF_renormalisation} and
Ref.~\cite{Reichert:2021cvs}.

The bounce action~\eqref{eq:bounce_action} is evaluated for the scalar field profile which solves
the corresponding equation of motion
\begin{equation}
      \frac{{\rm d}^2\sigma}{{\rm d}r^2}
    + \frac{2}{r} \frac{{\rm d}\sigma}{{\rm d}r}
    - \frac{1}{2} \frac{{\rm d}\log Z_\sigma}{{\rm d}\sigma}
      \left(\frac{{\rm d}\sigma}{{\rm d}r}\right)^2 =
    Z_\sigma \frac{{\rm d}\Veff(\sigma,T)}{{\rm d}\sigma}
    \; ,
\end{equation}
with boundary conditions
\begin{equation}
    \frac{{\rm d}\sigma (r=0,T)}{{\rm d}r} = 0 \; , \quad
    \lim_{r\to\infty} \sigma = 0 \; ,
\end{equation}
where $r$ is the radius of the nucleating bubble.
For a given temperature, we numerically obtain the critical bubble profile by employing a modified version of
the {\tt CosmoTransitions} package~\cite{Wainwright:2011kj}.

Next, let us introduce the relevant temperature scales.
To determine the onset of bubble nucleation, we demand
\begin{equation}
\label{eq:nucleation_criterion}
    \Gamma(\Tn) = H(\Tn)^4
    \;,
\end{equation}
which corresponds to the emergence of approximately one bubble per horizon.
The nucleation temperature $\Tn$, however, is not a reliable indicator of
the completion of the phase transition.
To this end, we define the percolation temperature $\Tp$ via the probability for a point in space to remain in
the false vacuum~\cite{Guth:1979bh,Guth:1981uk},
\begin{equation}
    P = \exp \left[-I(T)\right]
    \; .
\end{equation}
The exponent is given by~\cite{Ellis:2018mja}
\begin{equation}
\label{eq:I(T)}
    I(T) = \frac{4\pi}{3} \int_T^{\Tc}
      \frac{{\rm d} T'}{T'^4} \frac{\Gamma(T')}{H(T')}
      \biggl(\int_T^{T'}\!{\rm d}\Tilde{T} \frac{\vw }{H(\Tilde{T})}\biggr)^3
      \; ,
\end{equation}
where
$\vw$ is the bubble wall velocity.
Starting at the critical temperature $\Tc$, we track the evolution of $I(T)$ until
$I(\Tp) = 0.34$~\cite{Ellis:2018mja}.
This corresponds to $P \approx 0.7$, which we use as the criterion for successful percolation.
Therefore, all relevant parameters are evaluated at $\Tp$.
Inside the hadronic bubbles, the BSM scalar $\Phi$ rolls down to its potential minimum.
Therefore, a large amount of latent heat is involved, from which we conclude that
we can safely take $\vw = 1$.

Since the Universe is exponentially expanding during the $\chi\mathrm{PT}$,
we need to ensure that the volume of space in the true vacuum configuration indeed increases.
This imposes an additional constraint on $I(T)$ (cf., e.g., Ref.~\cite{Kierkla:2022odc}),
\begin{equation}
\label{eq:condition_false_vacuum_decreasing}
    \frac{1}{V_\mathrm{false}} \frac{{\rm d} V_\mathrm{false}}{{\rm d}t} =
    H(T) \left(3 + T \frac{{\rm d}I(T)}{{\rm d}T}\right) < 0
    \; ,
\end{equation}
where $V_\mathrm{false}$ is the volume which remains in the false vacuum.
This yields an upper bound on $\Ti$.

\begin{figure}[t]
    \centering
    \includegraphics[width=\columnwidth]{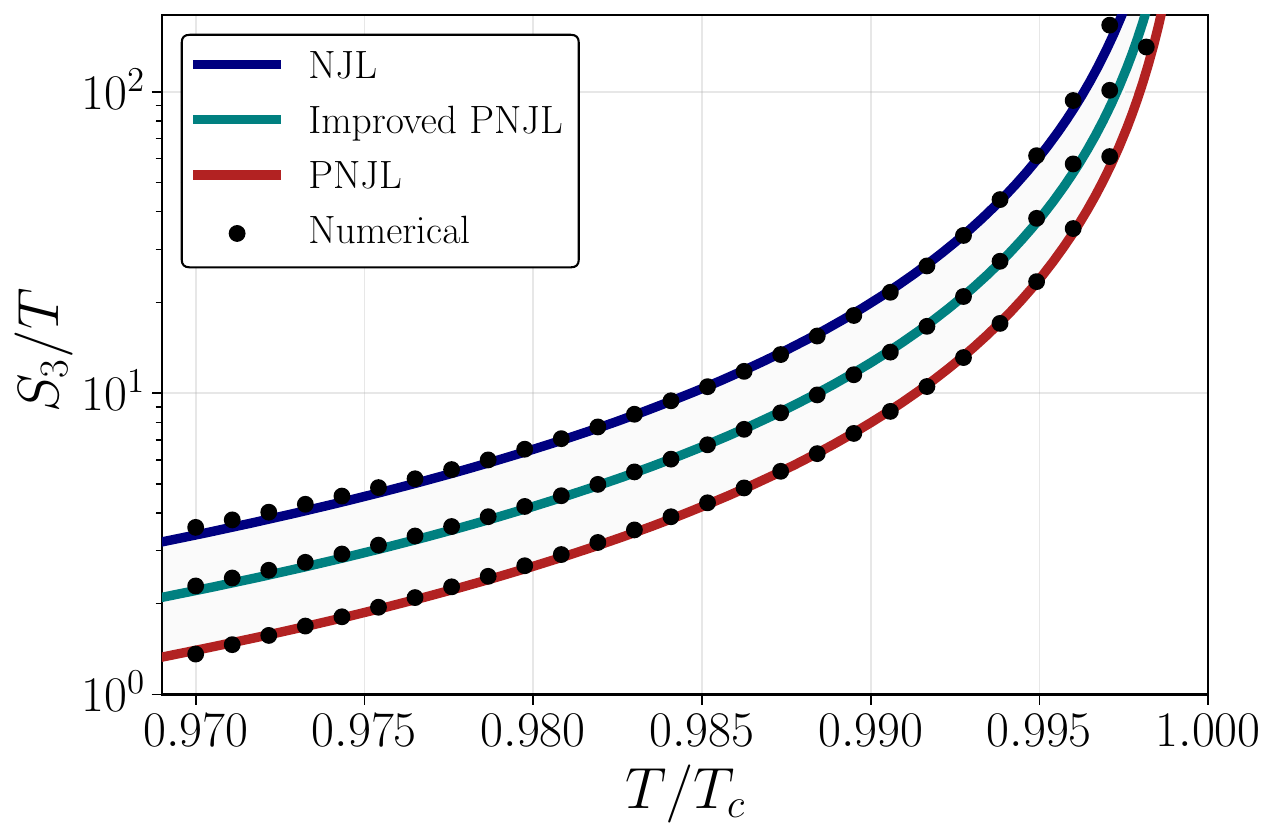}
    \caption{%
      Three-dimensional bounce action $S_3$ as a function of temperature.
      Numerical results obtained from {\tt CosmoTransitons} (dotted) are
      contrasted with the best fits from Eq.~\eqref{eq:fit_function} (colored lines).
      The fit parameters are found in Table~\ref{tab:best_fit}.}
    \label{fig:S3T_4D}
\end{figure}

The different QCD effective models share similar characteristics of the phase transition.
Figure~\ref{fig:S3T_4D} shows the evolution of $S_3/T$ within the (improved) (P)NJL model.
The results from our numerical evaluation (dotted) are contrasted with
the best fit as obtained with the parametrization from Ref.~\cite{Helmboldt:2019pan}:
\begin{equation}
\label{eq:fit_function}
    \frac{S_3 (T)}{T} \simeq b \Bigl(1-\frac{T}{\Tc}\Bigr)^{-\gamma}
    \; .
\end{equation}
The best fit values are listed in Table~\ref{tab:best_fit}.
These fits are used for the evaluation of the PT parameters.
For all three models, we observe that the bounce action drops rapidly with decreasing temperature. This indicates that the relevant dynamics take place close to $\Tc$. Overall, the Polyakov loop extended versions feature a rather steep slope compared to the NJL model. This already hints towards a decrease in the transition timescale by the inclusion of the Polyakov loop.
\begin{table}
\setlength\tabcolsep{7pt}
\begin{tabular}{|c|c|c|c|}
    \hline
     & \bf{NJL} & \bf{PNJL} & \bf{Impr.\ PNJL}\\
    \hline
    $b$ & $0.0118$ &  $0.0054$ & $0.0085$ \\
    \hline
    $\gamma$ & $1.614$ & $1.583$ & $1.585$ \\
    \hline
\end{tabular}
\caption{%
  Best fit values of the bounce action $S_3/T$ for different quark-based QCD effective models
  obtained with the parametrization
  in Eq.~\eqref{eq:fit_function}
  from Ref.~\cite{Helmboldt:2019pan}.}
\label{tab:best_fit}
\end{table}

\begin{figure}
    \centering
    \includegraphics[width=\columnwidth]{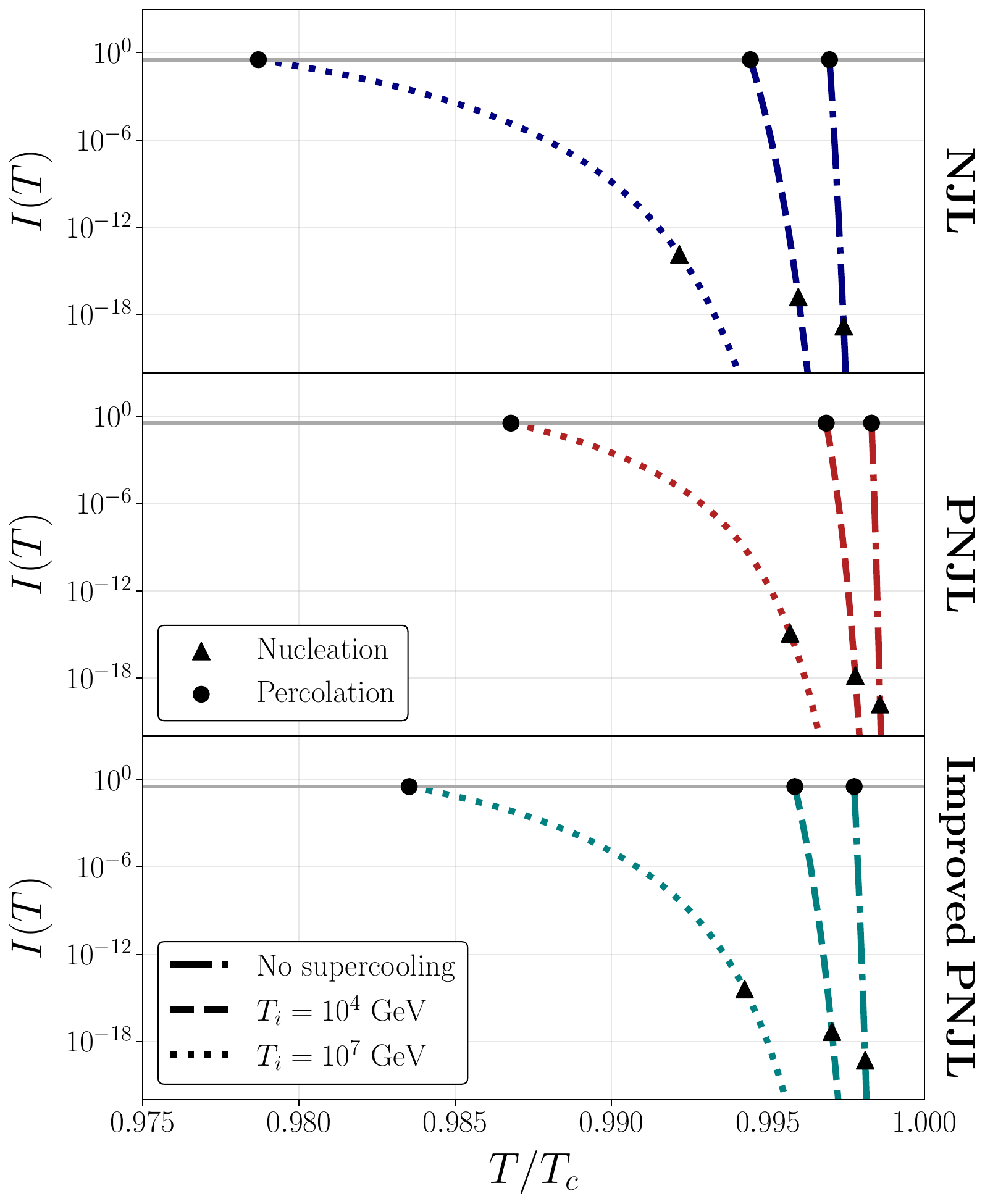}
    \caption{%
      Suppression of the false vacuum $I(T)$ via Eq.~\eqref{eq:I(T)}
      as a function of temperature.
      We employ three benchmark values for $\Ti$.
      The onset of nucleation according to Eq.~\eqref{eq:nucleation_criterion} is marked by the triangles, while the black dots denote $I(\Tp) = 0.34$~\cite{Ellis:2018mja},
      which is the criterion for successful percolation.
    }
    \label{fig:I_T_NJL}
\end{figure}
The corresponding exponential suppression of
the false vacuum $I(T)$ from Eq.~\eqref{eq:I(T)} is shown
in Fig.~\ref{fig:I_T_NJL} as a function of temperature.
The onset of bubble nucleation is indicated by the triangles, while the horizontal line marks $I(T) = 0.34$, thus successful percolation.
Here, we choose three example values for $\Ti$.
The corresponding benchmark without supercooling is less realistic in our setup and
rather resembles the dynamics of a dark sector chiral phase transition as, e.g.,
considered in Refs.~\cite{Helmboldt:2019pan,Reichert:2021cvs}.
However, we include this case as a reference to illustrate
the impact of the inflationary expansion.
We observe that with an earlier onset of thermal inflation both the
nucleation $\Tn$ and
percolation temperature $\Tp$ are decreased.
This is a consequence of the large expansion rate in a supercooling Universe.
Since the fraction of space in the true vacuum is suppressed by $H$,
a larger tunneling rate or conversely lower $\Tp$, is required to
render bubble nucleation efficient.

\begin{figure}
    \centering
    \includegraphics[width=\columnwidth]{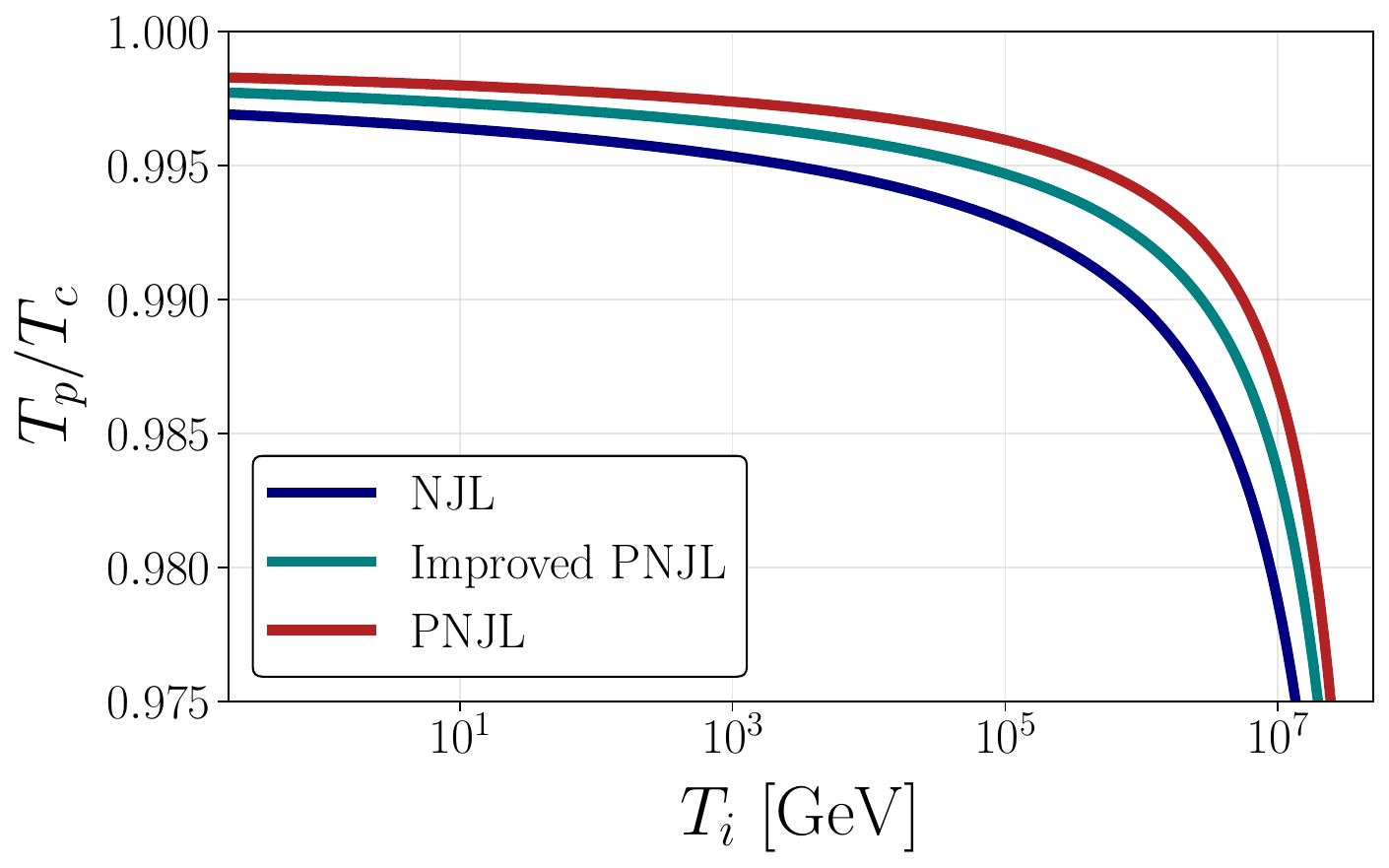}
    \caption{%
      Percolation temperature $\Tp$ as function of $\Ti$.
      For little supercooling, bubble percolation completes rapidly after the thermal bath passes $\Tc$.
      With increasing $\Ti$, the temperature scales drift apart.
    }
    \label{fig:Tp_Ti}
\end{figure}
To see this explicitly, we display the ratio of
the percolation and
critical temperature $\Tp/\Tc$ as a function of $\Ti$
in Fig.~\ref{fig:Tp_Ti}. While for little supercooling this ratio is close to one, the percolation temperature is considerably lowered with increasing $\Ti$.
This indicates an enlarged transition timescale in a supercooled Universe.

Based on the above results, we now focus on computing the parameters relevant for GW emission,
with particular focus on the impact of the supercooling period.

%%%%%%%%%%%%%%%%%%%%%%%%%%%%%%%%%%%%%%%%
%%%%%%%%%%%%%%%%%%%%%%%%%%%%%%%%%%%%%%%%
\section{Gravitational wave background}
\label{sec:GW_signal}
Anisotropies that were generated in the early Universe
via first-order phase transitions
can induce a characteristic observational relic:
a stochastic gravitational wave background.
We first introduce relevant quantities such as
the energy budget and
transition timescale before
discussing the efficiency coefficients and
the estimation of contributions of different GW sources within our setup.
To conclude, we show the predicted GW spectrum for
the different QCD effective models, depending on the amount of supercooling.
\\

\paragraph*{\bf Transition strength $\alpha$ and bubble wall speed $\vw$.}
The transition strength
of a cosmological phase transition reads~\cite{Caprini:2019egz,Hindmarsh:2017gnf}
\begin{align}
  \alpha(T) &\equiv \frac{1}{\rho_\mathrm{rad.}}
    \biggl(\Delta \Veff - \frac{T}{4} \Delta \frac{{\rm d}\Veff}{{\rm d}T}\biggr) \biggr|_{T = \Tp}
  \nn
  &\simeq \frac{\Delta \Veff}{\rho_\mathrm{rad.}(\TQCD)}
  \; ,
\end{align}
where $\Delta \Veff$ denotes the potential energy of the metastable vacuum.
As discussed in Sec.~\ref{sec:model}, we are interested in scenario~\ref{scenarioII} where quark condensation triggers the end of supercooling.
Therefore, we can define
the transition strength
in a model-independent fashion via the temperature where thermal inflation starts:
\begin{equation}
\label{eq:alpha_definition}
    \alpha \simeq \frac{\rho_\mathrm{rad.}(\Ti)}{\rho_\mathrm{rad.} (\TQCD)} =
    \left(\frac{\Ti}{\TQCD}\right)^4 \gg 1 \; .
\end{equation}
Thus, the latent heat which drives
the bubble expansion is extremely large and
we can assume
\begin{equation}
    \vw = 1
    \;,
\end{equation}
for the bubble wall velocity. \\

\paragraph*{\bf Inverse timescale $\beta / H$.}
In contrast to the energy budget of the transition, the inverse timescale is fully determined by QCD dynamics.
The false vacuum energy of the conformal SM can only be released in patches of space where the chiral symmetry is broken. Therefore, the duration of the transition is set by the nucleation rate of hadronic bubbles.

From the suppression of the false vacuum decay rate
\begin{equation}
  \Gamma(t) \propto e^{\beta t}
  \; ,
\end{equation}
with time $t$, one obtains
\begin{equation}
\label{eq:betaH_standard}
  \frac{\beta_\star}{H} =
  \Tp \frac{{\rm d}}{{\rm d}T} \frac{S_3(T)}{T} \Bigr|_{T = \Tp} \; ,
\end{equation}
for the inverse timescale of the transition.
All parameters with subscript $\star$ are computed at the time of percolation.
For our numerical evaluation, however, we employ an alternative definition of
$\beta_\star/H$ by introducing the average bubble radius at collision~\cite{Turner:1992tz,Enqvist:1991xw}
\begin{equation}
\label{eq:R_star}
    R_\star = \biggl[\Tp \int_{\Tp}^{\Tc}
      \frac{{\rm d}T'}{T'^2} \frac{\Gamma(T')}{H(T')} e^{-I(T')}\biggr]^{-\frac{1}{3}}
    \; .
\end{equation}
This method is widely used in the literature and found to be more robust against numerical instabilities compared to the standard definition in Eq.~\eqref{eq:betaH_standard}.
From Eq.~\eqref{eq:R_star}, the inverse timescale of the transition is calculated via 
\begin{equation}
\label{eq:beta}
    \beta_\star = \frac{\eta}{R_\star} \; ,
\end{equation}
where we again take $\vw = 1$ following the previous reasoning. The standard choice~\cite{Caprini:2019egz} for the proportionality factor is $\eta = (8\pi)^\frac{1}{3}$. However, recent simulations~\cite{Lewicki:2022pdb} have shown that $\eta \approx 5$ for strongly supercooled PTs.

\begin{figure*}
    \centering
    \includegraphics[width=\textwidth]{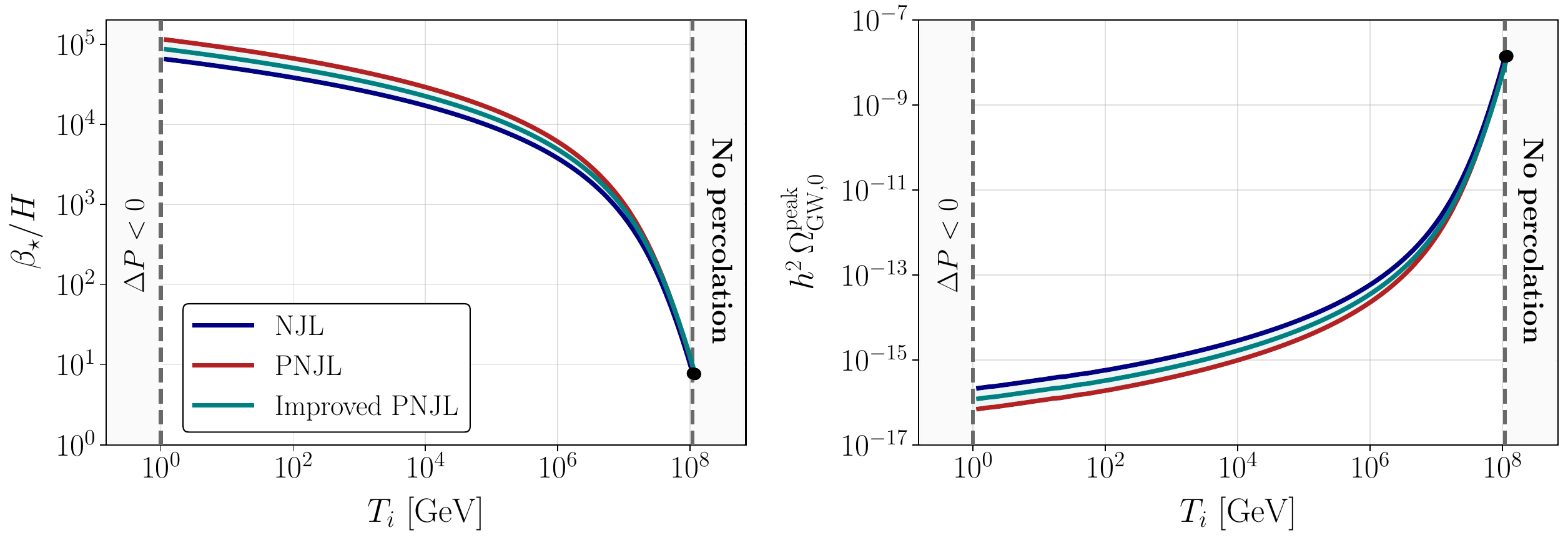}
    \caption{%
      Left:
      inverse timescale of the transition normalized to the Hubble parameter for different $\Ti$.
      The duration of the transition increases for longer supercooling periods.
      Right: GW peak amplitude as a function of $\Ti$, redshifted to today (indicated by the subscript $0$).
      As a consequence of the enlarged timescale, the GW amplitude is enhanced for large $\Ti$.
      In both plots, the shaded region on the left displays the parameter space where the false vacuum energy is too small to account for the friction exerted by the mass gain of the electroweak gauge bosons and the top quark [cf.\ Eq.~\eqref{eq:pressure_difference}].
      The shaded region on the right denotes the maximum
      $\Ti \approx 10^8~\GeV$ for which the true vacuum can efficiently expand
      [cf.\ Eq.~\eqref{eq:condition_false_vacuum_decreasing}].}
    \label{fig:beta_H_T_NJL}
\end{figure*}
The inverse timescale
$\beta_\star/H$ as a function of $\Ti$ is shown
in Fig.~\ref{fig:beta_H_T_NJL}~(left).
The region with no significant supercooling is comparable to studies of,
e.g., dark chiral phase transitions in QCD-like hidden sectors~\cite{Helmboldt:2019pan,Reichert:2021cvs}.
In agreement with the results therein, we find
$\beta_\star/H = \mathcal{O}\left(10^4-10^5\right)$ for all models.
Similar results are obtained for both chiral and
confinement/deconfinement phase transitions in holographic studies \cite{Morgante:2022zvc,Bigazzi:2020avc} of strongly coupled theories.
These values are orders of magnitude larger compared to the inverse timescale of cosmological phase transitions in nonstrongly interacting models, where typically
$\beta_\star /H = \mathcal{O}\left(100\right)$. With an increasing duration of thermal inflation, however, we observe that the inverse timescale clearly declines. To provide some analytical understanding, we insert
Eq.~\eqref{eq:fit_function} into
Eq.~\eqref{eq:betaH_standard}, which yields
\begin{equation}
    \frac{\beta_\star}{H} = \gamma \left(\frac{\Tc}{\Tp} - 1\right)^{-1} \frac{S_3(T)}{T}\; .
\end{equation}
From this expression, one can read off that in a supercooled Universe,
lowering the ratio $\Tp/\Tc$
(cf.\ Fig.~\ref{fig:Tp_Ti}) directly decreases $\beta_\star/H$.

This trend, however, terminates at a maximum value
$T_{i,\mathrm{max}}$, beyond which the inflationary temperature cannot be raised without violating
Eq.~\eqref{eq:condition_false_vacuum_decreasing}.
For larger $\Ti$, the expansion rate prevents the hadronic phase from expanding efficiently.
For all three QCD effective models, this corresponds to
$\beta_\star /H = \mathcal{O}(5-10)$ and
$T_{i,\mathrm{max}} = \mathcal{O}(10^8~\GeV)$, translating to approximately
$N_\mathrm{max} \simeq 20$ $e$-folds of thermal inflation.
If supercooling started above
$T_{i,\mathrm{max}}$, other mechanisms would have to be invoked to realize the exit from supercooling,
e.g.\ the destabilization of the false vacuum by the growth of quantum fluctuations~\cite{Lewicki:2021xku}.
This is however beyond the scope of this work and we therefore restrict ourselves to
$\Ti \leq T_{i,\mathrm{max}}$. Regarding the dependence on the employed QCD model, we find an $\mathcal{O}(1)$ deviation between the NJL and PNJL model in the low-$\Ti$ region, which shrinks as supercooling becomes more prominent.

The enhanced timescale has crucial consequences for the resulting GW spectrum which is
suppressed by~\cite{Caprini:2019egz,Lewicki:2020jiv,Hindmarsh:2017gnf,Hindmarsh:2015qta,Hindmarsh:2013xza,Giblin:2014qia}
\begin{equation}
    \Omega_\rmii{GW} \propto \left(\frac{H}{\beta_\star}\right)^n \; .
\end{equation}
As addressed in the next section,
the exponent $n = \{1,2\}$ depends on the GW source
which is based on the fact that fast bubble nucleation yields small radii at collision.
To obtain large spatial perturbations and therefore a strong GW signal,
the transition timescale, or $\Ti$, should be sufficiently large.
In Fig.~\ref{fig:beta_H_T_NJL}~(right),
we exemplify the impact of $\Ti$ on the GW peak amplitude from bubble collisions.
We observe that the spectrum is enhanced by $\sim 8$ orders of magnitude between the lower and upper limit of $\Ti$.

This is one of the main results of our work.
In addition to the large latent heat that is released, if QCD initiates the exit of supercooling, the duration of the transition is significantly increased.
As a consequence, the suppression of the GW amplitude is considerably weakened.
We now discuss different contributions to the GW spectrum.
\\

\paragraph*{\bf Gravitational wave sources.}
\label{sec:GW:source}
First-order phase transitions feature different sources of GWs.
In addition to gravitational radiation induced by the collision of bubbles of true vacuum, the interaction of the bubble wall with the surrounding plasma generates GWs from the propagation of sound waves and the formation of turbulences. Which source is the most prominent is determined by the friction between the bubble wall and the thermal bath. The pressure difference across the wall reads
\begin{equation}
\label{eq:pressure_difference}
    \Delta P = \Delta \Veff - P_{1\to 1} - P_{1\to N} \; ,
\end{equation}
where the first term denotes false vacuum energy. Although bubble nucleation is governed by QCD, we expect the new physics to quickly dominate the bubble expansion.
Therefore, the latent heat is primarily sourced by the new physics and
[cf.\ Eq.~\eqref{eq:Ti_definition}]
\begin{equation}
    \Delta \Veff \sim \Ti^4 \; .
\end{equation}
To successfully realize the exit from supercooling, this energy budget
has to withstand the inward pressure exerted by all particles in
the classically conformal SM.
To leading order (LO), the pressure is given by the second term in
Eq.~\eqref{eq:pressure_difference}.
This corresponds to friction from particles gaining a mass during the transition.
We have~\cite{Bodeker:2009qy}
\begin{equation}
\label{eq:LO_pressure}
    P_{1\to 1} \simeq \sum_i \frac{c_i k_i}{24} M_i^2 \Tp^2 \; ,
\end{equation}
where the sum is over all contributing particle species $i$,
$c_i = 1~(1/2)$ for bosons (fermions), and
$k_i$ denote
the massive degrees of freedom of the respective species.

The third term in
Eq.~\eqref{eq:pressure_difference} accounts for transition radiation at the bubble wall. There exist different results in the literature for this NLO contribution which differ in the scaling behavior of
the bubble wall Lorentz factor $\gamma$:
\begin{align}
\label{eq:friction1}
    P_{1\to N} &\simeq \gamma^{2} \sum_i k_i^{ } g_i^{ } \Tp^4
    \; ,\\[2mm]
    \label{eq:friction2}
    P_{1\to N} &\simeq \gamma \sum_{i} g_i^{ } M_{i}^{ } \Tp^3
    \; ,
\end{align}
given by
Refs.~\cite{Hoche:2020ysm,Gouttenoire:2021kjv}, respectively.
Here, all bosons $i$ contribute that couple to the bubble wall with coupling strength $g_i$.
The equilibrium $\gamma$ factor is then obtained by setting $\Delta P = 0$
\begin{equation}
\label{eq:gamma_eq}
    \gammaeq = \left(\frac{\Delta \Veff - P_{1\to 1}}{P_{1\to N}/\gamma^n}\right)^\frac{1}{n} \; ,
\end{equation}
where $n$ is the power of $\gamma$ appearing in
Eqs.~\eqref{eq:friction1} and \eqref{eq:friction2}, respectively.
At bubble collision, the Lorentz factor reads~\cite{Ellis:2019oqb}
\begin{align}
  \gamma_\star &= \min(\tilde{\gamma}_\star,\gammaeq)
  \; ,
  &{\rm with}
  &&
  \tilde{\gamma}_\star &\simeq \frac{2}{3} \frac{R_\star}{R_0}
    \; ,
\end{align}
where
$R_\star$ is given by Eq.~\eqref{eq:R_star}, and
$R_0$ is the initial radius of a nucleated hadronic bubble.
If $\gamma_\star = \gammaeq$, a terminal velocity is reached before collision, and plasma sources are the dominant source of GWs.
Whether such a steady state can be reached is
highly sensitive to the scale $\Ti$, the corresponding particle masses, as well as the gauge couplings.
This cannot be computed reliably within our model-independent framework.
Therefore, we restrict ourselves to a more qualitative discussion.

If supercooling commences well above the electroweak scale at $\Ti > T_\rmii{EW,SM} \gg \TQCD$, the vacuum energy is necessarily dominated by the new physics and we have
\begin{equation}
    \Delta \Veff \sim  M^4 \sim \Ti^4 \gg \Tp^4\; ,
\end{equation}
where $M$ is the gauge boson mass of the conformal sector. The relation between $\Delta \Veff$ and $M$ is a direct consequence of radiative symmetry breaking.
Because of the large hierarchy of scales, we may conclude that both
the LO and NLO friction contributions are strongly suppressed.
Hence, $\gammaeq$ is rendered large and we expect
$\gamma_\star = \Tilde{\gamma}_\star$, i.e.\ a runaway transition with bubble collisions
as the dominant source of GWs.
This has been verified, e.g., for the
$\UBL$~\cite{Ellis:2020nnr} and
$\SUX$~\cite{Kierkla:2022odc} extended SM.

If, on the other hand, thermal inflation sets in between the QCD and the EW scale, $\TEW > \Ti > \TQCD$, the scale hierarchy is less severe. Then the generation of GWs via sound waves might become efficient. Our model-independent ansatz does however not allow for precise statements since we lack information of the new physics.
A lower limit on $\Ti$ can be determined
by assuming that there are no masses above the electroweak scale.
Friction is then determined by the SM fields.
We therefore demand that the vacuum energy is sufficiently large to overcome
the LO pressure
$\Delta \Veff = P_{1\to 1}$~\eqref{eq:LO_pressure}
exerted by the top quark and the EW gauge bosons
i.e.\ $M_i \in \{m_t,\mZ,\mW\}$.
This gives
\begin{equation}
\label{eq:Ti_min}
    T_{i,\mathrm{min}} \approx 1~\GeV
    \;,
\end{equation}
as an approximate, model-independent lower bound on $\Ti$.
Otherwise, a negative net pressure would prohibit the broken EW phase to expand efficiently.

The behavior of the different GW sources in strongly supercooled PTs has been studied in
Ref.~\cite{Lewicki:2022pdb}.
For large transition strengths $\alpha \gg 1$, it was found that fluid dynamics play a negligible role since fluid shells propagate as relativistic shocks.
As a consequence,
sound-wave and collision-sourced GWs are equally suppressed and show a comparable spectrum.
In the case of a gauged SM extension,
the provided fit template is given by~\cite{Lewicki:2022pdb}
\begin{equation} \label{eq:GW_template}
    \Omega_{\rmii{GW},\star} =
      \left(\frac{H}{\beta_\star}\right)^2
      \left(\frac{\kappa_\mathrm{eff} \alpha}{1+\alpha}\right)^2 S_\mathrm{fit}(f)
      \;,
\end{equation}
with its spectral shape
\begin{equation}
    S_\mathrm{fit}(f) =
    A (a+b)^c \biggl(\frac{f}{f_p} \biggr)^a \biggr[
        b
      + a \biggl(\frac{f}{f_p} \biggr)^\frac{a+b}{c}
    \biggr]^{-c}
  \!\!
  \; ,
\end{equation}
for modes which are inside the horizon at collision. The parameters $A$, $a$, $b$, $c$, and the peak frequency $f_p$ are adopted from~\cite{Lewicki:2022pdb}. In addition, we set
$\kappa_\mathrm{eff} = 1$ due to the same spectral shape -- within errors -- of collision and sound wave sources. \\

\begin{figure*}
    \centering
    \includegraphics[width=\textwidth]{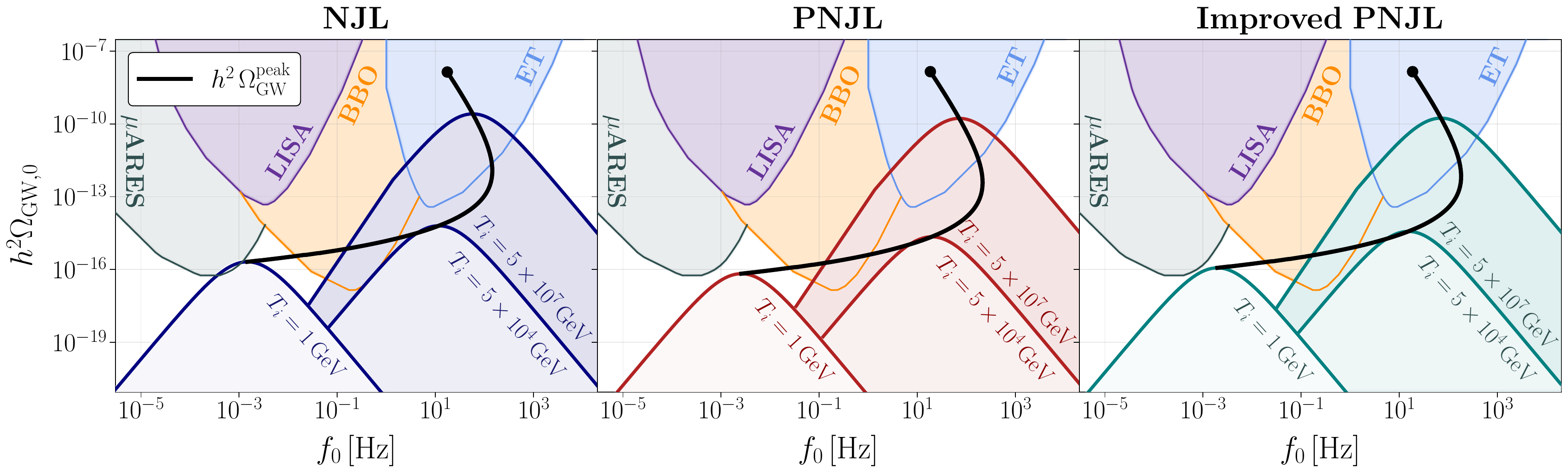}
    \caption{%
        Today's GW spectra from the QCD chiral phase transition in a supercooled Universe
        for three different QCD effective models:
        NJL (left),
        PNJL (middle), and
        improved PNJL (right).
        The benchmark spectra are computed via Eq.~\eqref{eq:GW_template} and correspond to different supercooling periods prior to the QCD scale. We impose $\Ti = 1~\GeV$,
        $\Ti = 5\times 10^4~\GeV$, and
        $\Ti = 5\times 10^7~\GeV$. In addition, we display the power-law integrated sensitivities of several future experiments. The black curve evolves the spectral peak with increasing amount of supercooling, and the black dot denotes the maximum $T_{i,\mathrm{max}} = \mathcal{O}(10^8~\GeV)$ which can be obtained without violating Eq.~\eqref{eq:condition_false_vacuum_decreasing}. From the enhancement of the transition timescale, the GW amplitude grows considerably with increasing $\Ti$, significantly improving the observational prospects.
        }
    \label{fig:GW_spectra}
\end{figure*}

\paragraph*{\bf Reheating period and redshift.}
To obtain predictions on the detectability of the $\chi$PT, we evolve the GW spectrum from the time of collision until today. To do so, we need to take into account that the Universe undergoes a period of reheating after the PT. In this stage, the fraction of false vacuum energy which has neither been converted into plasma motion nor gravitational radiation is transferred back to the heat bath. If reheating proceeds sufficiently slow, a phase of matter domination is induced where the BSM scalar $\Phi$ oscillates around its true minimum.
As a consequence, the GW abundance is diluted
$\Omega_\rmii{GW} \propto a_\star/a_\rmii{RH}$, where
$a_\rmii{RH}$ ($a_\star$) is the scale factor at the end of reheating (percolation).
The rate of energy transfer $\Gamma_\Phi$ from $\Phi$ to the SM is again highly model dependent, which makes a general conclusion hard to achieve.
In the following, we therefore assume $\Gamma_\Phi > H(\TQCD)$.
Then, reheating completes quickly and we have
\begin{equation}
    \TRH \approx \Ti
    \; ,
\end{equation}
which corresponds to a sudden decay of $\Phi$. For models with a finite reheating period, the resulting GW spectrum might be suppressed~\cite{Ellis:2020nnr}.

The redshifted GW spectrum and frequency read
\begin{align}
     h^2\Omega_{\rmii{GW},0} &= \Omega_{\rmii{GW},\star} \left(\frac{H_\star}{H_0}\right)^2 \left(\frac{a_\star}{a_0}\right)^4
     \nn &=
     1.67 \times 10^{-5}\left(\frac{100}{g_{\star}(\TRH)}\right)^\frac{1}{3}
     \Omega_{\rmii{GW},\star}
     \;, \\
     f_0 &= f_\star \frac{a_\rmii{RH}}{a_0}
     = f_\star \frac{T_0}{\TRH} \left(\frac{g_\star (T_0)}{g_\star(\TRH)}\right)^\frac{1}{3}
     \; ,
\end{align}
where the subscript $0$ ($\star$) denotes quantities evaluated today (at percolation).
\\

\paragraph*{\bf Observational prospects.}
Today's gravitational wave spectra as predicted by the (improved) (P)NJL model are
shown in Fig.~\ref{fig:GW_spectra}.
We depict three benchmark bubble collision spectra, varying
the duration of the supercooling period prior to $\csb$ and
evolve the spectral peak (solid, black).
For the spectral peak,
$\Ti$ is varied between the minimum and maximum value allowed by
Eqs.~\eqref{eq:Ti_min} and \eqref{eq:condition_false_vacuum_decreasing}, respectively.
The colored regions display the prospected power-law integrated sensitivities of several future observatories.
If there is an overlap between the spectra and the sensitivity curve, the corresponding signal is considered detectable.

For the benchmark at
$\Ti = 1~\GeV$, the predicted GW signal is strongly suppressed.
This is a consequence of the short timescale
$\beta_\star/H = \mathcal{O}(10^4-10^5)$ of the transition. For all models, the signal is barely visible by most future experiments. The most promising prediction is obtained within the NJL model, where the GW spectrum shows a slight overlap with $\mu$ARES~\cite{Sesana:2019vho}.

The second benchmark spectra correspond to an onset of thermal inflation at
$\Ti = 5 \times 10^4~\GeV$, hence well above the electroweak scale. The inverse timescale of the transition is considerably decreased, which leads to a boost of the GW amplitude, while the spectral peak moves to larger frequencies.  
Hence, we find the best detection prospects for the Big Bang Observer (BBO)~\cite{Crowder:2005nr}.
The shift of the peak frequency can be understood from the scaling relation
[cf.\ Eq.~\eqref{eq:Hubble_Ti}]
\begin{equation}
    f_\star \propto \frac{\beta_\star}{H(\Ti)} H(\Ti) \propto\frac{\beta_\star}{H(\Ti)} \Ti^2
    \; .
\end{equation}
From the redshift, we have a factor of
$\TRH^{-1} \simeq \Ti^{-1}$,
and
\begin{equation}
\label{eq:f0_scaling}
    f_0 \propto \frac{\beta_\star}{H(\Ti)} \Ti
    \; .
\end{equation}
Therefore, the peak frequency is determined by the interplay between increasing $\Ti$ and
the induced suppression of $\beta_\star/H$.
As a consequence, after a maximum, the peak frequency moves to smaller frequencies again, while
the amplitude is substantially enhanced.
This is the case for $\Ti \gtrsim 5 \times 10^6~\GeV$.
In this range, the suppression of $\beta_\star/H$ is more rapid than the increase of $\Ti$.
For this temperature range, the bubble collision spectra are pushed into the range of
ET~\cite{Sathyaprakash:2012jk,Punturo:2010zz}, as can be seen for the third benchmark point. Here, we imposed
$\Ti = 5 \times 10^7~\GeV$, which yields
$\beta_\star/H \approx 100$ (cf.\ Fig.~\ref{fig:beta_H_T_NJL}).
The low-frequency tail cuts into the sensitivity regions of the BBO.

The most optimistic scenario is denoted by the black dot and dictated by the maximum
$\Ti = \mathcal{O}(10^8~\GeV)$ allowed by
Eq.~\eqref{eq:condition_false_vacuum_decreasing}.
Beyond that limit, the expansion rate is too fast for the true vacuum to successfully expand.
Because of the steep spectral slope for small frequencies, we do not anticipate any observational prospects at
LISA~\cite{2017arXiv170200786A,Robson:2018ifk,LISACosmologyWorkingGroup:2022jok},
even for the most extreme scenarios.

%%%%%%%%%%%%%%%%%%%%%%%%%%%%%%%%%%%%%%%%
%%%%%%%%%%%%%%%%%%%%%%%%%%%%%%%%%%%%%%%%
\section{Conclusions}
\label{sec:conclusions}

In this work, we considered a mechanism which naturally arises in classically conformal SM extensions:
a period of thermal inflation which ends with the (first-order) QCD phase transition.
While the energy budget of such a transition is set by the false vacuum energy of the extended SM, bubble nucleation is governed by QCD.
To study these dynamics model independently,
we characterized the new physics merely by the temperature of the onset of supercooling.
To model chiral symmetry breaking with massless quarks, we employed three low-energy QCD effective theories:
the NJL model and
two Polyakov loop extended versions which incorporate the gluon dynamics.
These are constructed to reproduce certain properties of the QCD meson spectrum, and subsequently taken to the chiral limit.

Our main observation is that the inverse timescale of the transition, and therefore the strength of the GW signal, strongly depends on the background evolution. For little or no supercooling, the duration of the transition is short, leading to a large suppression of the GW amplitude.
This is consistent with previous findings in comparable studies of $\csb$ in dark QCD models~\cite{Helmboldt:2019pan,Reichert:2021cvs}.
We showed that in a supercooled Universe,
the resulting GW spectrum is considerably enhanced.
First, the exit from thermal inflation induces a large latent heat, thus a large transition strength $\alpha \gg 1$.
Furthermore, as a consequence of the rapid expansion of the Universe,
the timescale of the transition grows.
Hence, the GW amplitude is overall amplified. As the peak frequency is in addition shifted to larger values, we find the most promising observational prospects in the frequency regime governed by BBO and ET.
This is realized for an onset of thermal inflation well above the electroweak scale.

While our predictions are to a large extent model independent,
some model dependence remains.
The predicted timescale of the transition, and therefore the resulting GW amplitude, can differ by an $\mathcal{O}(1)$ factor between the different effective theories for small $\Ti$.
In the range where the transition is strong,
the model dependence becomes milder, hence our results are robust.
To model the strong dynamics more reliably, first-principle methods such as lattice techniques are required. This is left for future work.
In addition, the full
$\SU(6)_\rmii{L} \times \SU(6)_\rmii{R}$ symmetry of QCD should be included.
However, the impact of the inflationary period on the $\chi$PT will remain the same.
The enhancement of the transition timescale is a pure cosmological consequence, and
thus independent of the chosen model as long as it exhibits a first-order transition.
However, recent lattice studies~\cite{Cuteri:2021ikv} indicate that this requirement may not hold in the chiral limit, even for a large number of flavors.

Another intriguing aspect relegated to future work
is the impact of the Higgs field on the transition.
If quark condensation would first occur in the heavy quark sector,
the nonperturbative top quark Yukawa cannot be neglected~\cite{Bodeker:2021mcj}.
Then, more sophisticated techniques are required to treat the QCD + Higgs system.
Such an analysis cannot be model independent
since also new physics can alter the Higgs potential.

Our results are relevant for a wide class of models which feature similar dynamics.
Examples are
the scale invariant $\UBL$~\cite{Iso:2017uuu,Marzo:2018nov,Ellis:2020nnr,Jinno:2016knw} or
$\SUX$ extended SM~\cite{Prokopec:2018tnq,Kierkla:2022odc}, as well as
strongly coupled SM extensions~\cite{Baratella:2018pxi,vonHarling:2017yew}.
Such strongly supercooled phase transitions remain an exciting open topic.
At such low temperatures and potentially large field values,
the high-temperature expansion may be invalidated.
Thermal resummation necessary for infrared sensitive contributions
to the effective potential is therefore beyond
the current state-of-the-art formalism of
high-temperature dimensional reduction~\cite{Kajantie:1995dw,Croon:2020cgk,Ekstedt:2022bff}.
We will address in the future,
if this formalism is applicable even for such extremely
supercooled scenarios as discussed in this work.
Then, a precision analysis of specific conformal SM extensions could
scope the parameter space where the presented scenario is realizable.

Lastly, our work may also have implications in the context of dark sector model building.
Since the predicted signals from hidden chiral or confinement phase transitions are typically weak,
our work may improve their observational prospects.
%%%%%%%%%%%%%%%%%%%%%%%%%%%%%%%%%%%%%%%%
%%%%%%%%%%%%%%%%%%%%%%%%%%%%%%%%%%%%%%%%
\begin{acknowledgments}
The authors thank D.~B{\"o}deker,
M.~Buballa,
M.~Lewicki,
M.~Reichert, and
J.~Schaffner-Bielich
for fruitful discussions.
The authors acknowledge support by the Deutsche Forschungsgemeinschaft (DFG, German Research Foundation) through
the CRC-TR 211 `Strong-interaction matter under extreme conditions' --
project number 315477589 --
TRR 211.
\end{acknowledgments}
%%%%%%%%%%%%%%%%%%%%%%%%%%%%%%%%%%%%%%%%

%%%%%%%%%%%%%%%%%%%%%%%%%%%%%%%%%%%%%%%%
\appendix

\section{Wave function renormalization}
\label{sec:WF_renormalisation}

\begin{figure}
    \centering
    \includegraphics[width=\columnwidth]{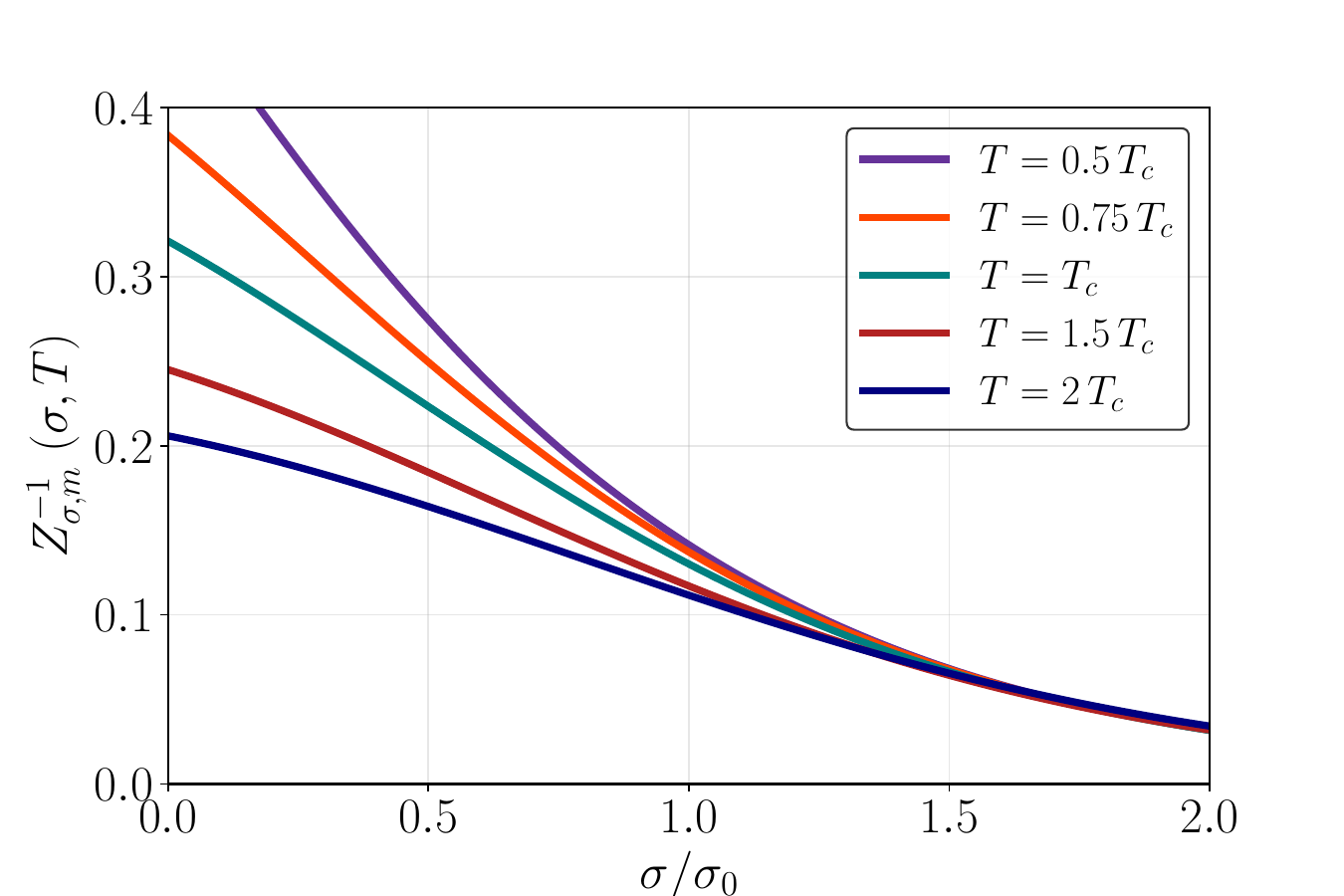}
    \caption{%
      Modified wave function renormalization factor $Z_{\sigma,m}^{-1}$,
      evaluated within the NJL model for different temperatures.}
    \label{fig:Z_sigma_NJL}
\end{figure}

Quark-based QCD effective models of NJL type are purely fermionic and
the quark condensate after bosonization, $\sigma$, does not propagate at tree level.
Its kinetic term is rendered only at loop level, hence
the bounce action~\eqref{eq:bounce_action}
is augmented by a wave function renormalization factor $Z_\sigma^{-1}.$
This section computes the wave function renormalization for the NJL model.
To this end, we largely follow~\cite{Reichert:2021cvs}, which
provides an instructive discussion of $Z_\sigma^{-1}$ for the PNJL model.

The wave function renormalization factor is computed from the $\sigma$ two-point correlation function.
From the NJL Lagrangian~\eqref{eq:NJL:lag:mfa} in the MFA~\cite{Holthausen:2013ota} and
the tree level potential~\eqref{eq:V0_sigma:NJL},
we identify the Feynman rules and obtain
two Feynman diagrams as radiative corrections at one-loop level for
the $\sigma$ two-point function
\begin{align}
\label{eq:sigma:1l}
  \Gamma_{\sigma\sigma}(q^0,\bf{q},\sigma) &\supset
  \parbox[c]{100pt}{%
  \includegraphics[width=0.4\columnwidth]{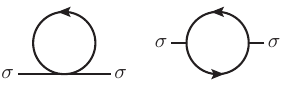}
  }
  \;,
\end{align}
where
solid lines are $\sigma$ fields and
directed lines are quarks.
The overall two-point function up to one-loop level is
\begin{align}
\label{eq:sigma_1PI}
  \Gamma_{\sigma\sigma}(q^0,\bf{q},\sigma) &=
    - \frac{3}{4 G}
    + \frac{3 G_\rmii{D}}{8 G^3} \sigma
    + \frac{G_\rmii{D}}{4G^2} \Nf \Nc I_\rmii{V}(\sigma)
  \nn
  &- \biggl(1-\frac{G_\rmii{D} \sigma}{4 G^2}\biggr)^2 \Nf \Nc I_\rmii{S}(q^0, \vec{q},\sigma)
  \; ,
\end{align}
where
$\Nf = \Nc = 3$,
$I_\rmii{V}(\sigma)$
and
$I_\rmii{S}(q^0,\vec{q},\sigma)$
correspond to
the loop integrals in Eq.~\eqref{eq:sigma:1l}, respectively.
At finite temperature, we have
$q^0 = i\omega_n^{\rmii{B}}$, where
$\omega_n^{\rmii{B}} = 2 n \pi T$ with
$n\in\mathbb{Z}$ denotes bosonic Matsubara frequencies.
Since $Z_\sigma^{-1}$ is computed from
\begin{equation}
  Z_\sigma^{-1} = -\frac{{\rm d}\Gamma_{\sigma\sigma}(q^0, \vec{q}, \sigma)}{{\rm d} \vec{q}^2}
    \biggr|_{\scriptsize
      \begin{aligned}
        q^0 &= 0
        \\[-1mm]
        \vec{q} &= 0
      \end{aligned}
      }\; ,
\end{equation}
it suffices to consider
the last integral
$\propto I_\rmii{S}(q^0, \bf{q},\sigma)$
in Eq.~\eqref{eq:sigma_1PI}
which gives rise to the only momentum dependence.
The relevant integral reads
\begin{align}
  I_\rmii{S}(i\omega_n^\rmii{B},&\vec{q}, \sigma)
    = \frac{1}{\Nc} \tr_\rmi{c}\, \Tint{\{P\}}^{\Lambda}
     \frac{\tr\, \left[
        (\msl{p} + M)
        (\msl{p} + \msl{q} + M)
      \right]}{
        \bigl[p^2 - M^2\bigr]
        \bigl[(p + q)^2 - M^2\bigr]
      }
    \;,
\end{align}
where
the curly brackets indicate the fermionic nature of thermal sums,
$\Tinti{\{P\}} = T \sum_{\omega_n^\rmii{F}} \int_{\vec{p}}$,
$M$ is given by Eq.~\eqref{eq:M:sigma},
$P = (p^0,\vec{p})$,
$p^0 = i \omega_n^{\rmii{F}}$, and
$\omega_n^{\rmii{F}} = (2n+1) \pi T$ are fermionic Matsubara frequencies.
\begin{figure}
    \centering
    \includegraphics[width=\columnwidth]{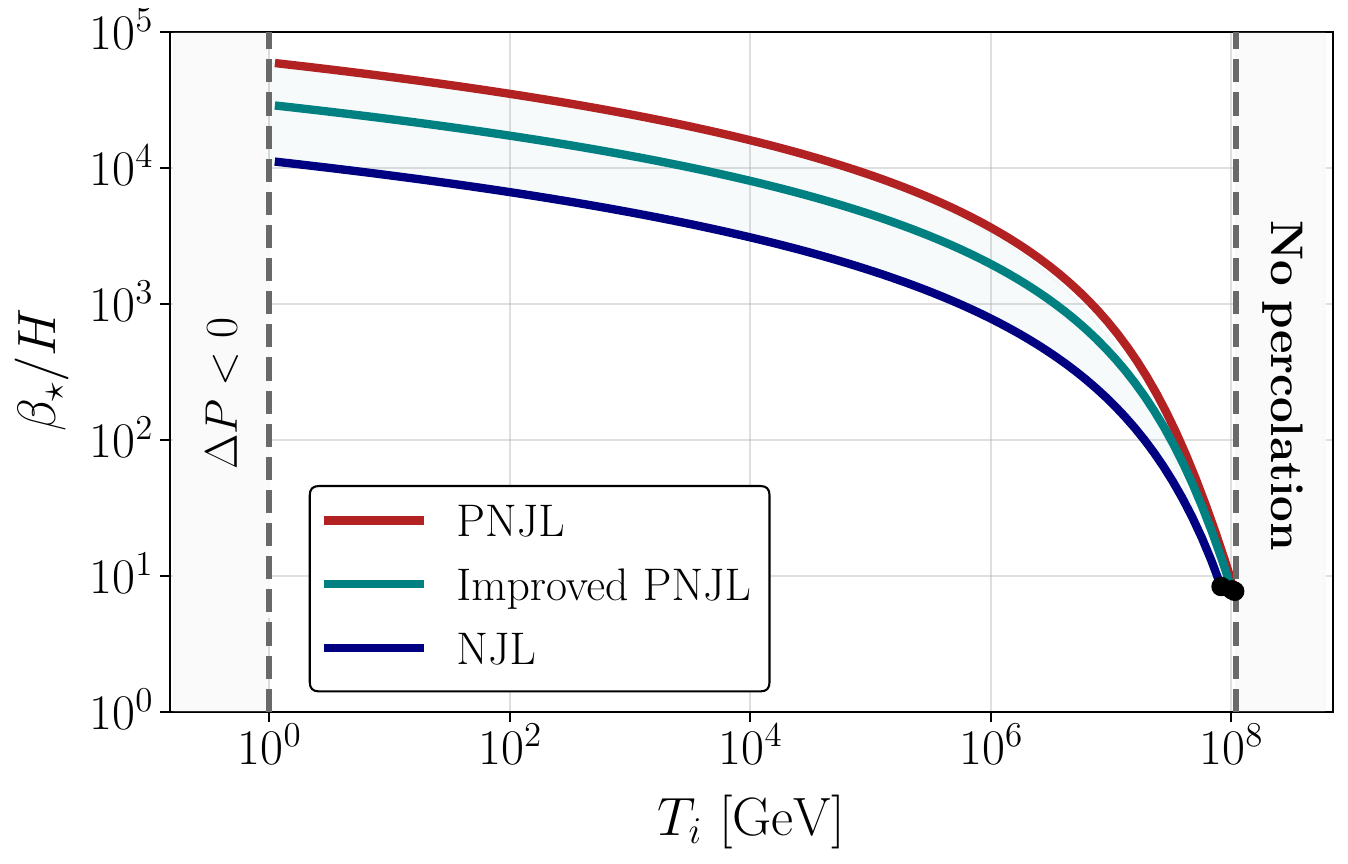}
    \caption{%
      Inverse timescale of the transition at the time of percolation
      normalized to the Hubble parameter for different $\Ti$
      computed within the 4D cutoff scheme;
      see Fig.~\ref{fig:beta_H_T_NJL} for the corresponding display
      within the 3D cutoff scheme.
      }
    \label{fig:betaH_4D}
\end{figure}
By decomposing~\cite{Reichert:2021cvs}
\begin{equation}
\label{eq:Z_decomposition}
    Z_\sigma^{-1} =
      -\biggl(1 - \frac{G_\rmii{D} \sigma}{4 G^2}\biggr)^2
      2 \Nf \Nc \left[I(0)+ 4 M^2 I'(0)\right]
    \;,
\end{equation}
we denote
\begin{align}
\label{eq:I_integral_WFR}
    I (i\omega_n^\rmii{B},&\mathbf{q},\sigma)
    \\
    &= \frac{1}{\Nc} \tr_\rmi{c}\,
    \Tint{\{P\}}^{\Lambda}
    \frac{1}{
        \bigl[p^2 - M^2\bigr]
        \bigl[(p + q)^2 - M^2\bigr]
    } \; ,
    \nn
    I(0)&= I (i \omega_n^\rmii{B} = 0, \mathbf{q} = 0, \sigma)
    \; ,\quad
    I'(0) = \frac{{\rm d} I}{{\rm d} \mathbf{q}^2}\biggr|_{\scriptsize
      \begin{aligned}
        \omega_n &= 0
        \\[-2mm]
        \vec{q} &= 0
      \end{aligned}
      }
    \;,
\end{align}
where
the integration $\Tintii{P}{\Lambda}$
denotes the strict three-dimensional momentum regularization
of the NJL model.
The strategy to evaluate the integral $I (i\omega_n^\rmii{B},\mathbf{q},\sigma)$
is the standard one~\cite{Kapusta:1979fh} employed in~\cite{Pisarski:1987wc,Parwani:1991gq}.
First, the sum over Matsubara modes
is converted into a twofold sum with a Kronecker delta function
$\delta(p_0) = T\int_{0}^{\beta} {\rm d}\tau \exp(i p_0 \tau)$.
The summation yields directly
\begin{align}
  T\sum_{\omega_n^\rmii{F}}
  &
  \frac{e^{i (\omega_n^\rmii{F} - i\mu) \tau}}{(\omega_n^\rmii{F} - i\mu)^2 + E^2}
  \\
  &=
  \frac{1}{2 E} \Bigl[
      f(E+\mu) e^{(\beta-\tau)E + \beta\mu}
    - f(E-\mu) e^{\tau E}
  \Bigr]
  \;,
  \nonumber
\end{align}
which we employ at zero chemical potential $\mu = 0$.
In this limit,
the Fermi-Dirac distribution is given by
\begin{equation}
    f(E) = \frac{1}{e^{\beta E} + 1}
    \;,
\end{equation}
where
$\beta = 1/T$.
For Polyakov loop extended models,
the fermionic distribution function $f(E)$
receives an $\ell$-dependent modification~\cite{Hansen:2006ee,Reichert:2021cvs}.
\begin{figure*}[t]
    \centering
    \includegraphics[width=0.9\textwidth]{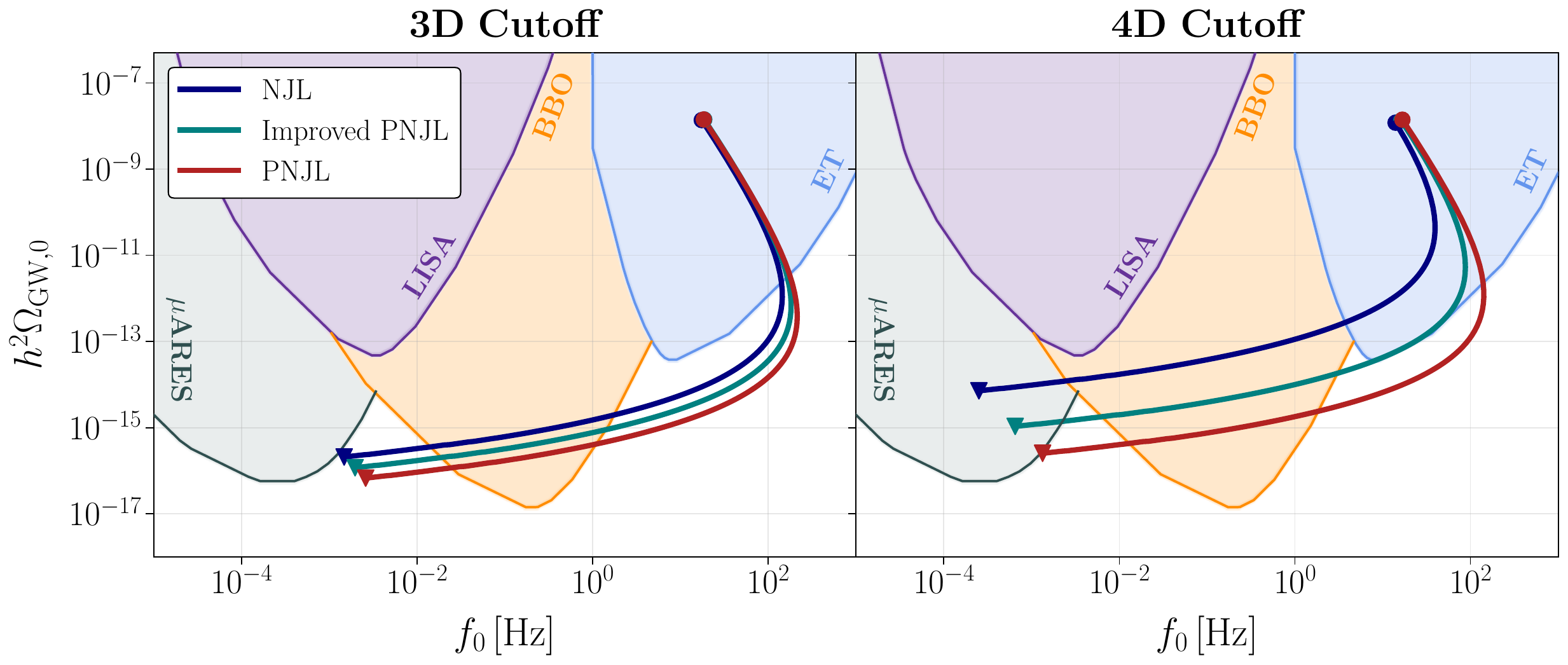}
    \caption{%
      Comparison of the evolution of the GW spectral peak between the 3D and 4D cutoff scheme, computed with Eq.~\eqref{eq:GW_template}.
      Different colors indicate the different quark-based QCD effective models.
      See also Fig.~\ref{fig:GW_spectra} for additional specifications.
    }
    \label{fig:Omega_GW_3D_4D}
\end{figure*}
The $\tau$ integral over the delta functions is straightforward,
leaving linear terms of the distribution functions in the numerator
with fractions of the three-energies
$E_p = \sqrt{p^2 + M^2}$.
We can write Eq.~\eqref{eq:I_integral_WFR} as
\begin{align}
  I(0, \mathbf{q},\sigma) = \int_{\vec{p}}^{\Lambda} \frac{1}{2 E_p E_{p+q}}
    &
    \biggl[\frac{f(E_p) - f(E_{p+q})}{E_{p} - E_{p+q}}
    \\
    +&\frac{1 - f(E_p) - f(E_{p+q}) }{E_{p} + E_{p+q}}\biggr]
    \; .
    \nonumber
\end{align}
By reinserting this expression into
Eq.~\eqref{eq:Z_decomposition},
we can compute $Z_\sigma^{-1}$.
Up to modifying the thermal distribution, this yields the same expression as given
in Ref.~\cite{Reichert:2021cvs}.
The final result, however, takes negative values for a large range of values, which is related to the breaking of Lorentz invariance in the 3D cutoff scheme.
To this end,
we adopt a modified wave function renormalization factor~\cite{Reichert:2021cvs},
{\em viz.},
\begin{align}
  Z_{\sigma,m}^{-1} &(\sigma,T) =
        \Bigl(1 - \frac{G_\rmii{D} \sigma}{4 G^2}\Bigr)^2 \frac{\Nc\Nf}{4\pi^2}
      \\ \times
      \biggl(
        &\int_0^\Lambda {\rm d}p \frac{p^2}{E_p^3}
        \Bigl[
          - 2f(E_p)
          + 2 E_p \frac{{\rm d}f(E_p)}{{\rm d} E_p}
          + 1\Bigr]
      \nn
      &+ M(\sigma)^2 \int_0^\Lambda {\rm d}p \frac{p^2}{E_p^5}
        \Bigl[
          6 f(E_p)
          - 2 E_p \frac{{\rm d}f(E_p)}{{\rm d}E_p}
          - 1\Bigr]
      \biggr) \; .
      \nonumber
\end{align}
Figure~\ref{fig:Z_sigma_NJL} shows
$Z_{\sigma,m}^{-1} (\sigma,T)$ for different temperatures around $\Tc$ as a function of the chiral condensate.
\section{Cutoff scheme dependence}
\label{sec:cutoff_scheme_dependence}
In the main part of this work, we have employed a 3D cutoff to regularize the NLO part of the NJL potential.
To estimate the robustness of our results, we now compare our predictions with the results we obtain in the 4D cutoff scheme, which is e.g.\ used in
Refs.~\cite{Helmboldt:2019pan,Holthausen:2013ota}.
The corresponding model parameters are listed in Table~\ref{tab:NJL_params_4D}.
\begin{table}
\setlength\tabcolsep{6pt}
\begin{tabular}{ |c|c|c| }
\hline
$\Lambda\left[\mathrm{MeV}\right]$ &
$G \Lambda^2$ &
$G_\rmii{D} \Lambda^5$ \\
\hline
$930$ & $4.44$ & $130.30$
\\
\hline
\end{tabular}\\\vspace{.2cm}
\begin{tabular}{|c|c|c|c|}
\hline
     &\bf{NJL} & \bf{PNJL} & \bf{Impr.\ PNJL} \\
     \hline
     $b$ & $0.223$ & $0.008$ & $0.028$ \\
     $\gamma$ &$1.609$ & $1.735$ & $1.739$ \\
     $\Tc~[{\rm MeV}]$ &$71.71$ & $121.77$ & $101.10$\\
     \hline
\end{tabular}
\caption{%
  Upper panel:
  NJL parameters which reproduce the QCD meson spectrum
  for the 4D cutoff scheme~\cite{Ametani:2015jla}.
  Lower panel:
  best fit values of the bounce action $S_3/T$ for different quark-based QCD effective models
  obtained with the parametrization
  in Eq.~\eqref{eq:fit_function}.
  }
\label{tab:NJL_params_4D}
\end{table}
With a 4D momentum cutoff, the integral for the vacuum energy
in Eq.~\eqref{eq:NJL_oneloop} evaluates to
\begin{align}
    \VNJL_{1,\rmii{4D}} (\bar\sigma) &=
    -\frac{\Nc \Nf }{(4\pi)^2}\Lambda^{4}\biggl[\ln\left(1 + \xi^2\right) \nn
    &\phantom{= -\frac{\Nc \Nf }{(4\pi)^2}\Lambda^{4}[]}- \xi^4 \ln\left(1 + \xi^{-2}\right) + \xi^2\biggr]
    \; ,
\end{align}
where $\xi = M/\Lambda$.
The thermal contributions to the effective potential are the same as in Eq.~\eqref{eq:NJL:1lT}.
Since these are naturally three dimensional,
the thermal integrals now cannot be treated with the same cutoff as the vacuum contribution.
Therefore, we take $\Lambda \to \infty$ for the thermal part.
Regarding the wave function renormalization, we use the expression of Ref.~\cite{Helmboldt:2019pan}.

We compute the bounce action as a function of temperature and
fit it to the parametrization~\eqref{eq:fit_function}.
The resulting fit parameters, together with the critical temperatures,
are found in Table~\ref{tab:NJL_params_4D}.
Since we take a larger temperature range into account for the fit,
we obtain slightly different fit parameters as
in Ref.~\cite{Helmboldt:2019pan} for the NJL and PNJL models.

The resulting inverse timescale is shown in Fig.~\ref{fig:betaH_4D} as a function of the temperature where thermal inflation starts. As expected, $\beta/H = \mathcal{O}(10^4)$ is large for small $\Ti$. When increasing the amount of supercooling prior to the QCD scale, the timescale of the transition becomes enhanced.
Thus we observe the same qualitative behavior as in Fig.~\ref{fig:beta_H_T_NJL}.
This is not surprising since the enlarged timescale is a consequence purely from the background evolution set by the conformally extended SM.

Some quantitative differences between the two cutoff schemes remain.
First, the 4D cutoff scheme generally predicts larger transition timescales
which is especially pronounced for small $\Ti$.
Here, the inverse timescale differs by about an order of magnitude,
for e.g.\ the NJL model, when compared to the value obtained in the 3D scheme.
The deviation becomes milder when increasing $\Ti$.
In the large-$\Ti$ range, both schemes agree reasonably well. Since this is the regime where a strong transition is expected, we consider our results robust. It is also worth noticing that the 4D cutoff scheme exhibits a considerably larger spread of the predicted timescales in between the different low-energy effective theories.
For short periods of thermal inflation, the different models span approximately
one order of magnitude between the minimal and maximal $\beta_\star/H$.
In the 3D scheme, we observe a much milder model dependence
(cf.\ Fig.~\ref{fig:beta_H_T_NJL}).

To illustrate the impact of the cutoff scheme on the observational prospects, we show the evolution of the peak amplitudes in Fig.~\ref{fig:Omega_GW_3D_4D}.
The two panels correspond to different cutoff schemes, while
colors indicate the chosen model.
The 4D cutoff approach produces overall stronger signals.
Already for the minimum amount of supercooling $T_i \approx 1~\GeV$, the predicted spectra show good overlap with the sensitivity region of $\mu$ARES. Such a feature is absent in the 3D cutoff scheme.

However, as already anticipated from the inverse timescale, the spectra obtained within the 3D scheme exhibit significantly less model dependence.
In the strong supercooling regime, on the other hand, the two schemes again agree reasonably well.
From that, we conclude to employ the 3D cutoff scheme, which is also more consistent at finite temperature.

%%%%%%%%%%%%%%%%%%%%%%%% Bibliography %%%%%
\bibliography{biblio}
%%%%%%%%%%%%%%%%%%%%%%%%%%%%%%%%%%%%%%%%
\end{document}